\documentclass[fleqn,usenatbib]{mnras}

\usepackage{newtxtext,newtxmath}

\usepackage[T1]{fontenc}

\usepackage{graphicx}
\usepackage{pdflscape}
\DeclareRobustCommand{\VAN}[3]{#2}
\let\VANthebibliography\thebibliography
\def\thebibliography{\DeclareRobustCommand{\VAN}[3]{##3}\VANthebibliography}

\DeclareMathOperator\erf{erf}
\usepackage[dvipsnames]{xcolor}
\usepackage[normalem]{ulem}

\title[Polarized emission from space particles]{Polarized microwave emission from space particles in the upper atmosphere of the Earth}

\author[J. L\'opez-Viejobueno et al.]{
Jennifer L\'opez-Viejobueno,$^{1,2}$\thanks{E-mail: jennilop@ucm.es}
Leire Beitia-Antero,$^{1,3}$
and Ana I. G\'omez de Castro$^{1,2}$\thanks{E-mail: aig@ucm.es}
\\
$^{1}$AEGORA Research Group, Facultad de Ciencias Matem\'aticas, Universidad Complutense de Madrid, Madrid, Spain\\
$^{2}$S.D. F\'isica de la Tierra y Astrof\'isica, Facultad de Ciencias Matem\'aticas, Universidad Complutense de Madrid, Madrid, Spain\\
$^{3}$Departamento de Estad\'istica e Investigaci\'on Operativa, Facultad de Ciencias Matem\'aticas, Universidad Complutense de Madrid, Madrid, Spain
}

\date{Accepted XXX. Received YYY; in original form ZZZ}

\pubyear{2023}

\begin{document}
\label{firstpage}
\pagerange{\pageref{firstpage}--\pageref{lastpage}}
\maketitle
\begin{abstract}
Tons of space particles enter the Earth atmosphere every year, being detected when they produce fireballs, meteor showers, or when they impact the Earth surface. Particle detection in the showers could also be attempted from space using satellites in low Earth orbit. Measuring the polarization would provide extra crucial information on the dominant alignment mechanisms and the properties of the meteor families. In this article, we evaluate the expected signal to aid in the design of space probes for this purpose. We have used the \textsc{RADMC-3D} code to simulate the polarized microwave emission of aligned dust particles with different compositions: silicates, carbonates and irons. We have assumed a constant spatial particle density distribution of 0.22~cm$^{-3}$, based on particle density measurements carried during meteor showers. Four different grain size distributions with power indices ranging from $-3.5$ to $-2.0$ and dust particles with radius ranging from 0.01~$\mathrm{\mu}$m to 1~cm have been considered for the simulations.
Silicates and carbonates align their minor axis with the direction of the solar radiation field; during the flight time into the Earth atmosphere, iron grains get oriented with the Earth's magnetic field depending on their size. Alignment direction is reflected in the $Q$-Stokes parameter and in the polarization variation along the orbit. Polarization depends on the composition and on the size distribution of the particles. The simulations show that some specific particle populations might be detectable even with a small probe equipped with high sensitivity, photon-counting microwave detectors operating in low Earth orbit.
\end{abstract}


\section{Introduction}

The observation of the Earth from space is a rich source of information of the global properties of the planet and its interaction with the spatial environment. It also provides important clues to understand the story of the Earth in the Solar System and serves as a baseline for the search of Earth-like exoplanets. 
Since tons of extraterrestrial material plunge into the Earth's atmosphere over a year, Earth observation supplies information about solar system bodies too. Although the mass input is uncertain and depends on the technique used, an average ranging from 5 to 50 tons per day is estimated \citep{Plane2012, Rojas2021}.
Most of these estimates of extraterrestrial material influx rely upon the measurements carried out by the networks monitoring short-duration fireball events and meteor showers \citep{Halliday1996, Bland2012, Howie2017, Jenniskens2018, Mane2021} and by some spatially limited ground based meteorite searches in deserted areas, either in hot deserts \citep{Bland1996, Hutzler2016} or in Antarctica \citep{Evatt2020, Rojas2021}.
Unfortunately, a global space based survey is still missing.

Near Earth object (NEO) forecast is becoming more popular and useful because these meteoroid showers might increase the risk of collision and damage on low Earth orbit spacecraft surface. For instance, \citet{Grun1985, Divine1993} and \citet{ Moorhead2017} developed an interplanetary dust model to obtain sporadic meteoroid flux, and some spatial agencies made meteoroid environment models: NASA’s Meteoroid Engineering Model (MEM) \citep{McNamara2004} and ESA’s Interplanetary Meteoroid Environment Model (IMEM) \citep{Soja2019}.

Most of these infalling particles are leftovers and debris from comets and shattered asteroids in near Earth orbit. The principal sources of this cosmic dust influx are Jupiter-family
comets (with a mass contribution of about 80--85 per cent), the Asteroid belt, Halley-type comets, and the Oort-Cloud comets \citep{Nesvorny2010, Carrillo2016}.
Should this material be observed, important clues on the composition of the NEOs would be attained that will complete our current understanding of these bodies achieved by the characterization of asteroids through their reflectivity \citep[see, e.g.,][]{Parker2008, Waszczak2013, Waszczak2015}. Also, a better analysis on the impact of space dust in the evolution of atmospheric reddening along the history of the Earth would be made feasible. 

The thermal emission from space dust is expected to be polarized due to the alignment of the grains. 
Numerous mechanisms have been advanced to produce this alignment, especially in the context of interstellar dust (see \citet{Andersson2015} for an overview of the field). Radiative alignment torques (RATs) are the best studied mechanism. Since the first suggestions on helical grain alignment by RATs  (\citet{Dolginov1976}, passing by the  
numerical studies by \citet{Draine_Weingartener1996, Draine_Weingartener1997} that proved the high efficiency of grain acceleration under radiation 
to the fundamental analytical models that bear the RAT alignment theory \citep{LH07}. Other significant alignment mechanisms that may be relevant in near-Earth orbit are 
mechanical torques by collisions with atmospheric gas \citep{Gold1952a, LH07b} or the Earth magnetic field; in the presence of a magnetic field, both radiative or mechanical torques can lead grains to be aligned with it under certain conditions. Thus, depending on the dust properties and the detailed environmental conditions, the axis of alignment will be: the direction of the radiation field ($k$-alignment), the direction of the magnetic field ($B$-alignment) or the direction of grain velocity ($v$-alignment).
Hence, the orientation with respect to the field lines and the orientation with respect to a potential observer in orbit will provide an additional mean to differentiate grain properties.

In this paper, we study the grain alignment mechanisms in the Earth's upper atmosphere, and  compute for the first time
the expected microwave signal of these infalling particles and its polarization,
in order to assess its detectability.
In Sect.~\ref{main_parameters}, we present a thorough review of the observational constraints on the
properties of the Earth-interplanetary medium interface. In Sect.~\ref{setup}, we present
our model for the radiative transfer calculations that have been performed with the
Monte Carlo code \textsc{RADMC-3D}\footnote{\url{https://www.ita.uni-heidelberg.de/~dullemond/software/radmc-3d/}} (version 2.0, \citealt{Dullemond2012}). 
In Sect.~\ref{results}, we analyse the properties of the predicted
polarization maps and in Sect.~\ref{discussion} we discuss the potential
detectability of this polarization with the small microwave probe being designed by the 
MARTINLARA (Millimeter wave Array at Room Temperature for INstruments in LEO Altitude Radio
Astronomy) consortium\footnote{MARTINLARA's web project: \url{https://martinlara3.webnode.es/}}. 
We conclude in Sect.~\ref{conclusions} with a brief summary of the main results.

\section{Particles at the Earth-space interface}  \label{main_parameters}

Modelling the interface between the upper layers of the atmosphere
and the interplanetary medium is very challenging since there
are many underlying uncertainties that arise from the lack of
systematic observations. For predicting the polarization from
infalling dust particles, we need to impose several constraints
on the dust properties, but also on the Earth environment. In
this section, we first review the current knowledge
on the properties of infalling dust particles: their
size distribution and composition 
(Sect.~\ref{properties}), their spatial distribution
(Sect.~\ref{spatial_distribution}), and their expected alignment based on RAT
theory (Sect.~\ref{alignment}).

\subsection{Dust grain properties and size distribution}\label{properties}

The flux of extraterrestrial material falling onto the Earth's surface is measured by the fireball monitoring networks \citep[see, e.g.,][]{Howie2017} but also by ground based searches on well defined search areas usually in the desert \citep{Bland1996, Hutzler2016} or in Antarctica \citep{Evatt2020}. The size distribution follows well the histogram put forward by \citet{Hughes1994} (see its fig. 20) and mathematically, it is correctly described by a power law,
\begin{equation}
\begin{centering}
    \frac{\mathrm{d}n(a)}{\mathrm{d}a} = n_0  a^{\alpha}
\end{centering}
\label{distribution}
\end{equation}
where $n_0$ is the scale factor, $a$ is the radius of the grain assumed to be spherical, and $\alpha$ is the index of the power law. As shown in \citet{Hughes1994}, the particle infall flux is divided into four regions: interplanetary dust particles (IDPs) and micrometeorites, meteors, fireballs, and crater-forming objects. Notice that the cometary and asteroid curves flatten at low diameters, making these small particles 
difficult to detect with the current facilities. Indeed, ground-based measurements provide scant information about the size distribution of particles entering the Earth's atmosphere at the low mass end. Nevertheless, \citet{Carrillo2015} show that the small dust grains ($<5$~$\mathrm{\mu}$m) might constitute a significant portion which is also expected 
since the disruption of large particles is expected to increase the number density of small particles \citep{Hoang2019}.

Several space missions have provided estimates of $\alpha$ for interplanetary dust. For instance, the SP-2 experiment carried by the Vega spacecraft observed comet 1P/Halley and determined
that the size distribution index does not adjust to only one curve, but it ranges from $-3.4$ to $-1.5$ for different intervals of mass \citep{Mazets1986}. Also, the spacecraft Giotto obtained measurements of
$\alpha = -2.6$ from the nucleus of comet Halley \citep{Fulle2000}. The Stardust mission estimated a value of $\alpha = -2.72$ from the size of the notches produced by the projectiles emanating from comet 81P/Wild2 and colliding with the device \citep{Horz2006}. The Rosetta spacecraft, orbiting around 67P/Churyumov-Gerasimenko comet, also provided observations for various samples with $\alpha \in [-3.8, -2.6]$ \citep{Merouane2017}; \citet{Ueda2017} considered an index of $-3.5$ for asteroids. Thus, we have opted to carry out a grid of simulations with fiducial values for $\alpha = (-3.5, -3.0, -2.5, -2.0)$. 

The upper and lower bounds of the size distribution are also ill-determined \citep{Guttler2019}. Regarding the aforementioned histogram in \citet{Hughes1994}, and considering that the bodies of interest are cometary and asteroid fragments and smaller, we have set a grain size range from $0.01~\mathrm{\mu}$m to 1~cm. The lower cutoff 
is set based on our estimates that the radiation produced (consequently, polarization) by grains smaller than $0.01~\mathrm{\mu}$m is weak at microwave wavelengths compared to the Earth background radiation. The upper cut-off is set to allow for the small particles resulting from the meteoroids break-up that are numerous enough to contribute to the polarization of the signal.
 
The most abundant materials in meteoritic dust are amorphous silicate (olivine, MgFeSiO$_4$), graphite, and metallic iron thus, these are the materials considered for this work. 
Silicate and graphite refractive indexes have been computed following \citet{Draine2003}\footnote{The refractive indices are publicly available for download at \url{https://www.astro.princeton.edu/~draine/dust/dust.diel.html}}. Since graphite is an anisotropic material, the refractive index depends on the orientation of the crystal, so we have followed the standard 1/3--2/3 approximation \citep{Draine&Malhotra} to obtain a scalar value. Iron complex refractive index data are from \citet[][Part II, Subpart I]{Palik1991} and are extended to the microwave range by applying a Drude model extrapolation with the parameters in \citet{Ordal1988}. Grain material density $\rho_{\rm g}$ is considered to be of 3.30~g\,cm$^{-3}$ for silicates, 2.24~g\,cm$^{-3}$ for carbonates, and 
7.87~g\,cm$^{-3}$ for metallic irons.

\subsection{Vertical distribution of space dust}\label{spatial_distribution}

The vertical distribution of space dust across the layers of the Earth's
atmosphere depends strongly on the dynamics and composition of the infalling particle shower; for instance, \citet{Mane2021} found that particles accumulated in the mesosphere and lower thermosphere during the Quadrantids meteor showers and, however, 
they gathered in the upper stratospheric layers for the Gamma Velids, Alpha Crucids or Delta Cancrids meteor shower in 2009. 

Little is known about the mass distribution above $\sim 130$~km height (very close to the limiting Karman line located at 100~km). Above this height, the interaction with the Earth's atmosphere is smooth and particle showers are difficult to detect from the ground \citep[see, e.g.,][]{Howie2017, Mane2021}. 
Below $\sim 130$~km, the number density of particles from meteor showers, $N(h)$, is roughly inversely proportional to the square of the height, that is, $N(h) = N(0)h^{\beta}$ with $\beta = -2$. For instance, data based on the atmospheric extinction \citep{Mane2013} by \citet{Mane2021} results in $\beta = -1.86$. At higher heights, since no fragmentation at the atmosphere is yet happening, the spatial particle density should be rather constant apart from the tidal effects associated with their trajectory and the entry angle in the atmosphere.

In the absence of more detailed information, we have derived the total particle density at 130~km from \citet{Mane2021} and we have assumed that the density above this height remains constant up. The most precise fitting curve during the Quadrantis and Gamma Velids showers in 2009 January 4 was:
\begin{equation}
  \log  N(h) = 3.27 -1.86 \log h
    \label{AND_h}
\end{equation}
with $N(h)$ given in particles (of all considered sizes) per cubic centimeter and $h$ in kilometers above the Earth's surface. Hence, for $h \sim 130$~km and above, the expected particle density is $N(130$~km$) \simeq 0.22$~cm$^{-3}$.

\subsection{Dust grain alignment}\label{alignment}
Grains in the Earth vicinity
may be aligned by various mechanisms. In general, the strongest interaction is with the solar radiation
field, which is anisotropic, resulting in a net radiative torque (RAT) and grain alignment \citep[see, e.g.,][]{LH07}. This radiative torque is given by
\begin{equation}
    \Gamma_{\rm rad} = \frac{\gamma \overline{u} a^2 \overline{\lambda}}{2}Q_{\rm \Gamma}
\end{equation}
where $\gamma$ is the anisotropy degree of the radiation field (set to unity), $Q_{\rm \Gamma}$ is the RAT efficiency (see below), and $\overline{u}$ and $\overline{\lambda}$ are the mean energy density of the solar radiation field and mean wavelength, defined as
\begin{equation}
    \overline{u}=\int u_{\rm \lambda}\,\mathrm{d}\lambda,
\end{equation}
\begin{equation}
    \overline{\lambda}=\frac{\int \lambda u_{\rm \lambda}\,\mathrm{d}\lambda}{\overline{u}}.
\end{equation}

The magnitude of the RAT efficiency can be approximated by

\begin{equation}
    Q_{\rm \Gamma} \simeq 0.4\left(\frac{\lambda}{1.8a}\right)^{\eta}
    \label{Q_RAT}
\end{equation}
where $\eta = 0$ for $\lambda\lesssim 1.8a$ and $\eta=-3$ for $\lambda>1.8a$.

Then, the timescale for RAT precession, $t_{\rm k}$, is
\begin{equation}
    t_{\rm k} = \frac{2\pi}{\overline{u} a^2 \overline{\lambda} \gamma \overline{Q}_{\rm \Gamma} / I \omega}
\end{equation}
being $I=8\pi\rho_{\rm g} a^5/15$ the inertia moment of the grain, $\omega = \sqrt{2k_BT/I}$ the grain angular velocity, with $k_B$ the Boltzmann constant and with $T$ the grain temperature,
and $\overline{Q}_{\rm \Gamma}$ the RAT efficiency average

\begin{equation}
    \overline{Q_{\rm \Gamma}}=\frac{\int Q_{\rm \Gamma} \lambda u_{\rm \lambda}\,\mathrm{d}\lambda}{\int \lambda u_{\rm \lambda}\,\mathrm{d}\lambda}.
\end{equation}

The radiation pressure from the Sun results in an alignment of the dust particles in near Earth orbit.
However, this alignment is bound to be disrupted at the atmospheric entry by collisions with atmospheric particles. 

\begin{figure}
    \centering
    \includegraphics[width=88 mm]{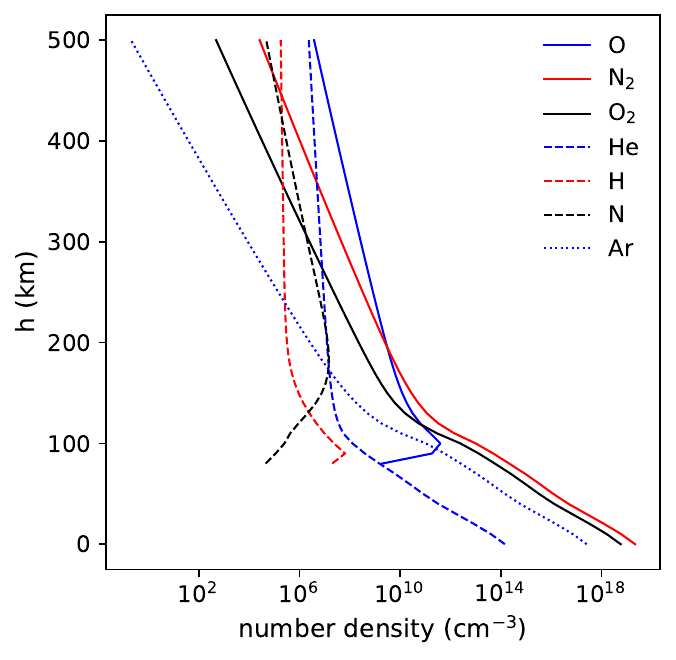}
    \includegraphics[width=88 mm]{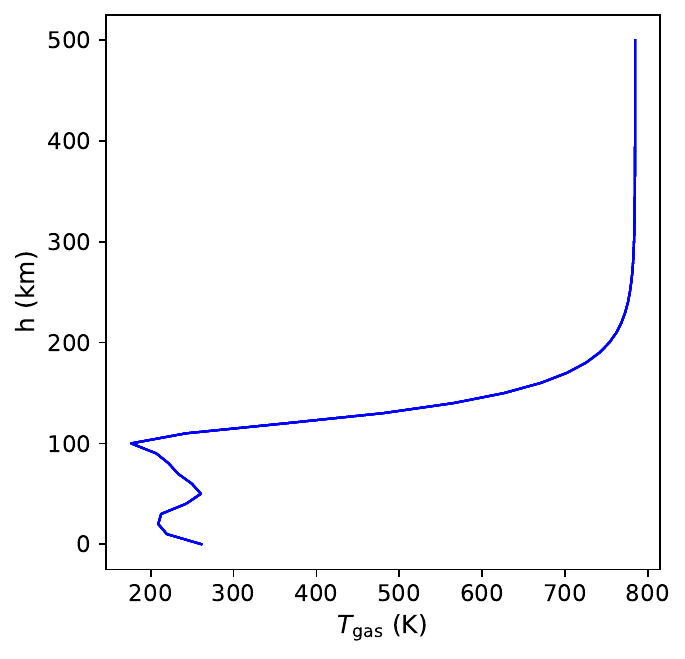}
    \caption{Number density of the neutral atoms and molecules of the atmosphere (\textit{top panel}) and 
     temperature of the gas atmosphere at the layers of interest (\textit{bottom panel}). Data are extracted from NRLMSISE-00 model
      at South Pole on August 4, 2020.}
    \label{fig:atmosphere}
\end{figure}

The dominant damping mechanism is the collision  with neutral particles (atoms and molecules) in the Earth's atmosphere. 
To estimate the critical height at which grain-neutral particle collisions damp any previous grain alignment,
we have used the dimensionless damping coefficient $F_n$ \citep{Draine_Lazarian1998}:

\begin{equation}
\begin{split}
    F_n = & \sum_{Z_g}f(Z_g)\sum_n \frac{n_n}{n_{\rm N_2}}\left(\frac{m_n}{m_{\rm N_2}}\right)^{1/2} \\
    & \times [\exp(-Z_g^2\epsilon_n ^2)+|Z_g|\epsilon_n \pi^{1/2}\erf(|Z_g|\epsilon_n)]
\end{split}
\end{equation}
where $f(Z_g)$ is the probability that a grain has a charge $Z_ge$, 
$m_{\rm N_2}$ and $n_{\rm N_2}$ are the mass and number density of N$_2$ in the atmosphere
respectively, and $m_n$ and $n_n$ are the mass and number density of the molecular and atomic 
abundances of the neutral species $n$ of the atmosphere. For our calculations, we have obtained
the atmospheric abundances from the NRLMSISE-00 model
\footnote{\url{https://kauai.ccmc.gsfc.nasa.gov/instantrun/nrlmsis/}} for the most abundant species 
up to 500~km, which are H, He, O, N, Ar, O$_{\rm 2}$, and N$_{\rm 2}$, (see Fig.~\ref{fig:atmosphere}). Finally,
the term $\epsilon_n$ is related to the polarization of the considered atom or molecule and is given by
\begin{equation}
    \epsilon_n^2\equiv \frac{e^2 \alpha_n}{2a^4 kT_{\rm gas}}
\end{equation}
where $\alpha_n$ it the polarizability of neutral species and $T_{\rm gas}$ is the temperature profile of the gas, 
also provided in Fig.~\ref{fig:atmosphere}. As a first approximation, we have considered
the dust particles to be completely neutral, so $Z_{g} = 0$ and $f(0) = 1$.

Hence, for neutral dust grains the damping time $t_{\rm damping}$  is given by
\begin{equation}
    t_{\rm damping} = \tau_{\rm N_2}/F_n
    \label{t_damping}
\end{equation}
where $\tau_{\rm N_2}$ is included only as a factor of normalization that is
given by \citep{Draine_Lazarian1998}:

\begin{equation}
    \tau_{\rm N_2} = 3I\left(\frac{\pi m_{\rm N_2}}{8k_BT_{\rm gas}}\right)^{1/2}\frac{1}{2\pi n_{\rm N_2}m_{\rm N_2}a^4}.
\end{equation}

\begin{figure}
    \centering
    \includegraphics[width=78 mm]{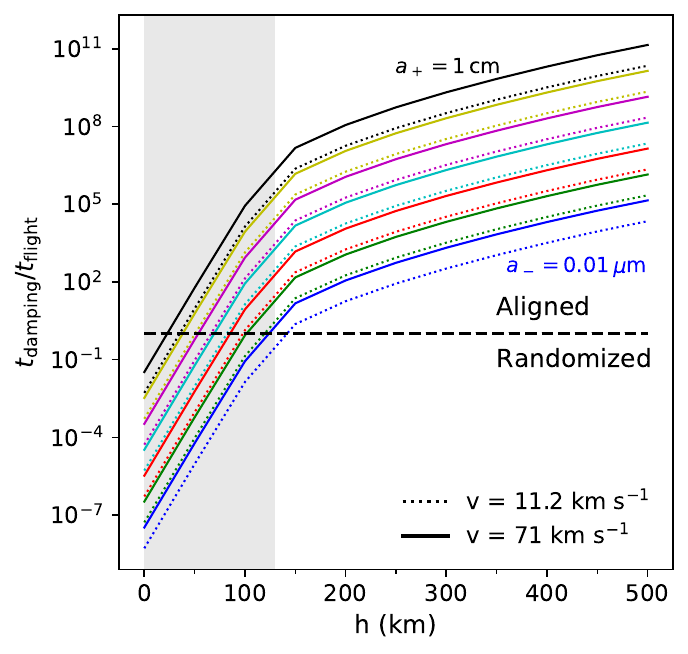}
    \includegraphics[width=78 mm]{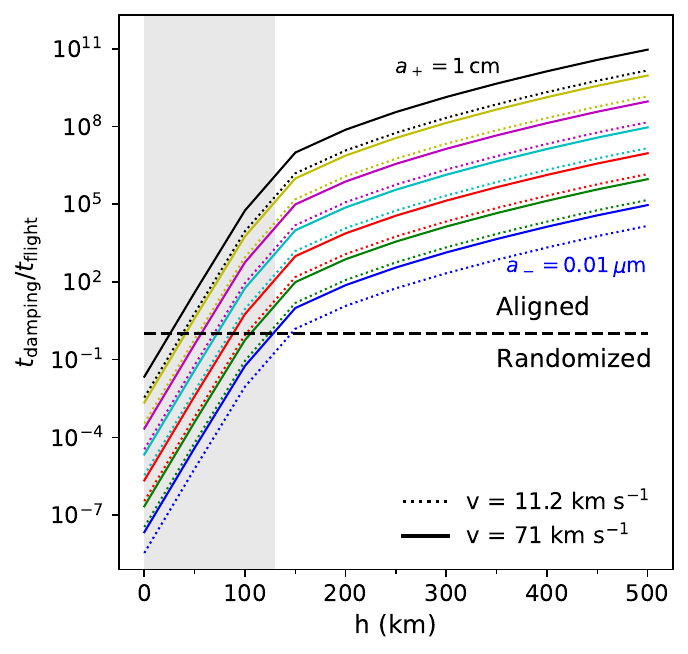}
    \includegraphics[width=78 mm]{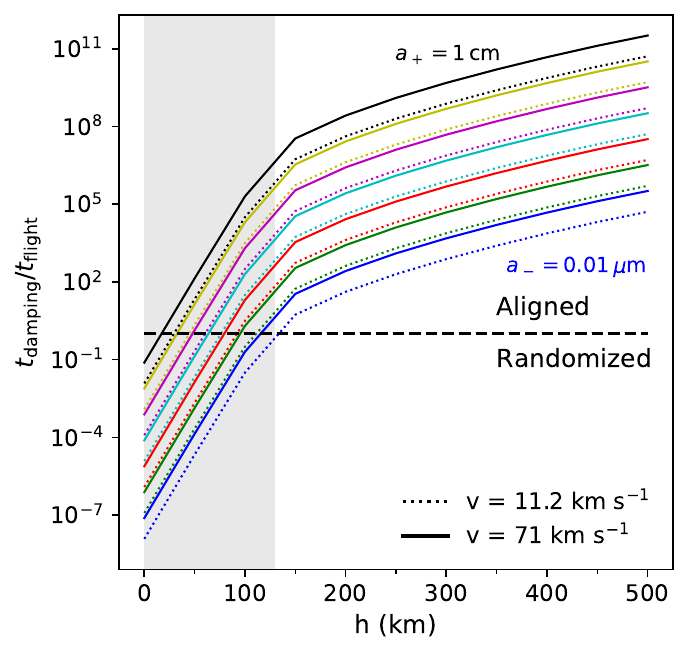}    
    \caption{Disruption of alignment for silicates (top panel), graphites (middle panel) and iron (lower panel)
    by grain collisions with the neutral gas as a function of height. The 
    shaded region corresponds to $h < 130$~km that limits
    the alignment of the smaller-sized particles.}
    \label{fig:alignment_disruption}
\end{figure}

Any initial alignment will be lost if the collision timescale is smaller than 
the flight time, $t_{\rm damping} < t_{\rm flight}$. 
The flight time is defined as the time
it takes for the particle to cross the atmosphere, and depends strongly on the orbital velocity of the space body captured by the Earth's gravitational field. 
For the calculation of the flight time, we have considered that the travelled space is twice 50~km (suggested size for the beam of the space probe) and that the velocity of the dust grains is in the range between the free-fall velocity (11.2~km\,s$^{-1}$)
and $\sim 71$~km\,s$^{-1}$, which roughly corresponds to the velocity of the Leonids, the fastest of the regular meteor showers.
Recent empirical evaluations have resulted on a similar range for the velocity of the particles falling on the Earth \citep{Carrillo2015,Carrillo2020}.

Ratios $t_{\rm damping}/t_{\rm flight}$ are displayed in Fig.~\ref{fig:alignment_disruption} for the three grain compositions.
As shown in the figure, any initial RAT-alignment is lost below 130~km by collisions with the dense atmospheric gas. This value is also the height at which most meteoroids start to break up \citep{Koten2004}. 
For this reason, the bottom boundary of the simulation domain will be set at 130~km (see below).

During their infall, dust grains are also spinning, acquiring a magnetic moment $\mu$ due to the Barnett effect (spontaneously magnetization of an uncharged body when it rotates) \citep{Barnett1915}. 
This magnetic moment $\mu$ interacts with the Earth's magnetic field $B$ causing a Larmor precession around the magnetic field direction. The precession rate, $t_{\rm B}$, is given by
\begin{equation}
    t_{\rm B} = \frac{2\pi I\omega}{\mu B}
\end{equation}
where the magnetic moment $\mu$ for paramagnetic grains is defined as
\begin{equation}
    \mu=\frac{\chi(0)V\hbar}{g\mu_B}\omega
\end{equation}
and $V$ is the volume of the grain, $\hbar=h/2\pi$ with $h$ the Planck constant, $g\approx 2$ is the gyromagnetic ratio, and $\chi(0)$ is the magnetic susceptibility at zero frequency. For silicates, $\chi(0)$ is given by Curie's law \citep{Morrish2001}:
\begin{equation}
    \chi(0)=4.2\times10^{-2}f_{\rm p}\left(\frac{T}{15~\mathrm{K}}\right)^{-1}
\end{equation}
with $f_{\rm p}$ the fraction of atoms in the grain that are paramagnetic, taken as 1/7. For iron grains, if they are smaller than 0.03~$\mathrm{\mu}$m a value of $\chi(0)=3.3$ is taken, otherwise $\chi(0)\approx 12$ \citep{DL99}.

For carbonaceous grains, following the analysis of \citet{Weingartner2006}, we consider 
as dominant the Barnett magnetic moment of hydrogenated carbonaceous grains:
\begin{equation}
    \mu({\rm carb})=1.5\times10^{-23}\left(\frac{T}{15~\mathrm{K}}\right)^{-1}\left(\frac{\omega}{10^5~\mathrm{s^{-1}}}\right)\left(\frac{a}{0.1~\mathrm{\mu m}}\right)^{3}.
\end{equation}

\begin{figure}
    \centering
    \includegraphics[width=88 mm]{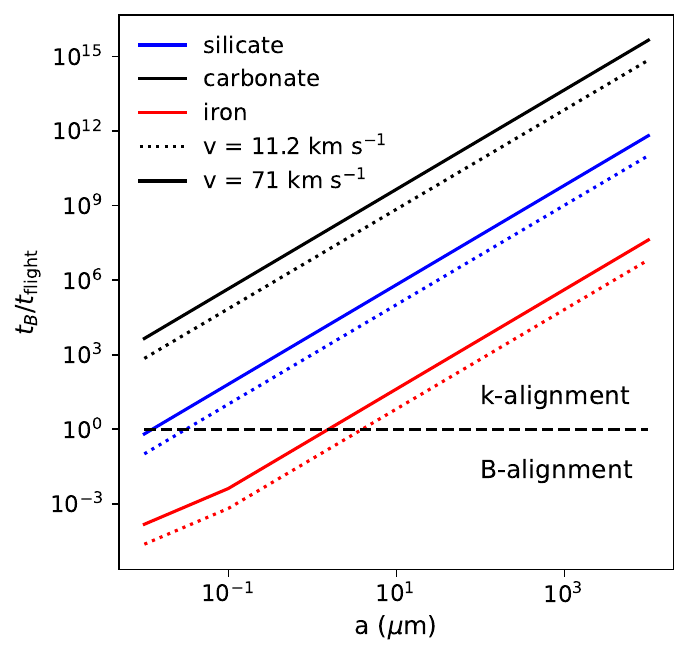}
    \caption{Ratio of $B$-alignment timescale over flight time for the species considered in
    this work. Assuming an initial $k$-alignment with the solar radiation field, two regions
    are distinguished in this plot: for $t_{\rm B}/t_{\rm flight} > 1$, particles do not have time to interact
    with the magnetic field and remain aligned with the radiation field, while for
    $t_{\rm B}/t_{\rm flight} < 1$ they become
    $B$-aligned if $t_{\rm B} < t_{\rm k}$.
    Travelled space is considered in stretches of 100~km.}
    \label{fig:t_B/t_flight}
\end{figure}

Magnetic alignment ($B$-alignment) is dominant if $t_{\rm B} < t_{\rm k}$,
otherwise the grain will align with the direction of the radiation field ($k$-alignment). 
Magnetic alignment depends on the strength of the Earth magnetic field that decreases
with distance as $1/r^2$. The dominant alignment mechanism and hence, the orientation of the polarization vector, will not only
depend on the grain composition, but also on the location of the dust grain in the magnetosphere. 
Moreover, the timescale for magnetic alignment has to be smaller than the flight time. 
If $t_{\rm B} < t_{\rm k}$ and $t_{\rm B} < t_{\rm flight}$, the timescale for the alignment of the dust grains with the Earth
magnetic field will be small enough for grains under the beam of the space probe. However,
if $t_{\rm B} < t_{\rm k}$ and $t_{\rm B} > t_{\rm flight}$, there will not be time enough for the dust grain to align with the B-field during its path under the telescope beam.
In Fig.~\ref{fig:t_B/t_flight} we represent the ratio $t_{\rm B}/t_{\rm flight}$ as a function of particle size
for the three dust families (silicates, graphites and iron) to discern the relevance of the $B$-alignment.
We see that silicates might become $B$-aligned for sizes smaller than $\sim$0.03~$\mathrm{\mu}$m 
and iron grains smaller than $\sim$4~$\mathrm{\mu}$m are $B$-aligned; carbonates are not $B$-aligned. 

We have checked that for all cases when $t_{\rm B}<t_{\rm flight}$ is achieved, also the required condition $t_{\rm B}<t_{\rm k}$ is obeyed.
For these calculations, the dipole model of the Earth magnetic field has been used \citep{bookGeomag}, having a Earth magnetic field magnitude ranging between 0.5 and 0.6~G. 
 
Charged carbonate grains may also be aligned by other mechanism in the presence of environmental magnetic fields.  
The motion of the grains with velocity $v$ through the field induces and electric field ($\textbf{E}$) that may result in an alignment of the grains with their long axis in the same direction than $\textbf{B}$ \citep{Lazarian2020}.

\begin{figure}
    \centering
    \includegraphics[width= 80mm]{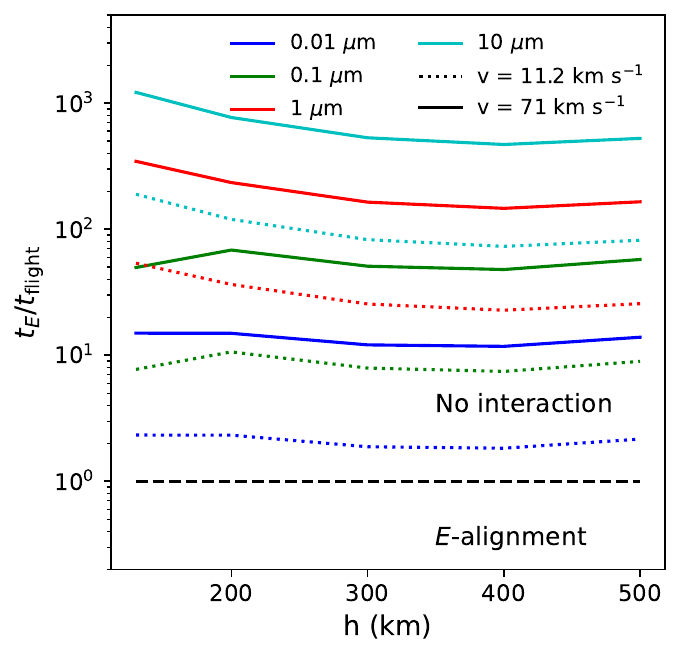}
    \caption{Timescale comparison between the precession around the induced electric field $t_E$ and the flight time $t_{\rm flight}$. $t_E$ is inefficient compared to $t_{\rm flight}$ and no $E$-alignment is produced.}
    \label{fig:tE_tflight}
\end{figure}

The electric precession timescale is given by,
\begin{equation}
    t_E = \frac{2\pi c I \omega}{p v B}
\end{equation}
with $c$ the speed of light and $p$ the grain electric moment that depends on the charge distribution of the grain. Calculations of grain charge are summarized in Appendix \ref{charge}.
To evaluate the relevance of this process, we compare the flight time of the grains with the precession timescale, see Fig.~\ref{fig:tE_tflight}. From the inspection of the figure, 
it becomes evident that even the smallest grains have not time enough to align with the induced electric field during their fall and that $E$-alignment is not relevant for this study.

To sum up, we will have aligned dust particles above 130~km 
(Fig.~\ref{fig:alignment_disruption}) and depending on composition and 
grain size, they will be aligned with the solar radiation
field or with the Earth
magnetic field (Fig.~\ref{fig:t_B/t_flight}).

\section{The radiative transfer model} \label{setup}
Our main goal is to compute the predicted polarization signal
arising from dust falling onto the Earth. For that purpose, we used the
radiative transfer code \textsc{RADMC-3D} that computes the emission
produced by dust grains for a given composition and
spatial distribution.

\begin{figure*}
    \centering
    \includegraphics[width = .95\textwidth]{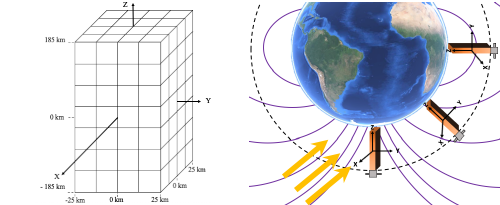}
    \caption{{\it Left:} Simulation grid and location of the centre of the reference system. We note that the computational domain in the vertical direction 
    ranges from $-185$~km to +185~km, that correspond to heights
    $h = 500$~km and $h = 130$~km respectively. {\it Right:} 
    Artistic representation of the orbit around the Earth. The three depicted positions are the ones
    simulated with \textsc{RADMC-3D} that correspond to three different relative alignments with line of sight
    (A: South pole -- parallel, C: equator -- perpendicular, B: intermediate -- 45$^{\rm o}$)
     The reference system is anchored in the satellite with $Z$ axis always pointing in the nadir direction and $X$ axis always pointing to the Sun (gold arrows); the computational domain for the Monte Carlo simulations is sketched in orange.}
    \label{fig:dust_satellite_conf}
\end{figure*}

For the simulations, we have defined an evenly-spaced 3D Cartesian grid with a resolution of 1~km in each direction with 50$\times$50$\times$370 cells. This results in a 50~km $\times$ 50~km grid projected onto the Earth surface; in the vertical direction, the grid starts at an altitude of 130~km and goes to a height of 500~km; the lower limit corresponds to the critical value below which dust grains are not aligned due to collision with the atmosphere
(see Sect.~\ref{alignment} for the detailed calculations), while the upper limit corresponds to a reference low Earth orbit (LEO).
The origin of our coordinate system is placed at the centre of the grid
(see Fig.~\ref{fig:dust_satellite_conf}) with the $Z$ axis pointing to nadir from the satellite, the $X$ axis pointing to the Sun and $Y$ defined following a positively right-handed system. 
This configuration is shown in Fig.~\ref{fig:dust_satellite_conf} for three different locations of the satellite.

One of the main parameters of our model is the dust density distribution. Due to the inherent uncertainties
on size distribution and composition (see Sect.~\ref{properties}), we opted for running four sets
of simulations for different values of the power law index $\alpha = (-3.5, -3.0, -2.5, -2.0)$
and for three dust families: olivine, graphites, and metallic irons. On each case, we computed the corresponding
grain opacities using the python version of the Bohren and Huffman code under the Mie theory assumption (BHMIE code) \citep{Bohren&Huffman} available in the
\textsc{RADMC-3D} package and using the optical properties provided in Sect.~\ref{properties}.

For setting the dust density value of each simulation cell, we recall our 
continuity assumption 
of dust particle number for $h > 130$~km, being $N = 0.22$~cm$^{-3}$ 
(Sect.~\ref{spatial_distribution}). This value can be used for the computation of the scale factor $n_{0}$ involved in Eq.~\ref{distribution} and depends on $\alpha$; as a
reference, the derived values of $n_{0}$ are shown in Table~\ref{tab:n0}.

\begin{table}
\caption{Values of the scale factor $n_0$ for the size distributions
considered in this work.}
\label{tab:n0}
\centering
\footnotesize
\begin{tabular}{lcccc}
\hline
\hline
$\alpha$ & $-3.5$ & $-3.0$ & $-2.5$ & $-2.0$ \\
\hline
$n_0$ & 5.44 $\times 10^{-16}$ & 4.35 $\times 10^{-13}$ & 3.26 $\times 10^{-10}$ & 2.18
$\times 10^{-7}$ \\
\hline
\end{tabular}
\end{table}

On the other hand, for setting a model in \textsc{RADMC-3D}
we need to provide discrete dust families instead of the continuous size distribution given
by Eq.~\ref{distribution}. Therefore, we have considered ten particle sizes logarithmically
spaced in the range 0.01~$\mathrm{\mu}$m--1~cm, resulting in bins centred at 
$a_1 = 0.02~\mathrm{\mu}$m, 
$a_2 = 0.09~\mathrm{\mu}$m, 
$a_3 = 0.34~\mathrm{\mu}$m, 
$a_4 = 1.36~\mathrm{\mu}$m, 
$a_5 = 5.42~\mathrm{\mu}$m, 
$a_6 = 21.58~\mathrm{\mu}$m, 
$a_7 = 85.90~\mathrm{\mu}$m, 
$a_8 = 341.98~\mathrm{\mu}$m, 
$a_9 = 1\,361.46~\mathrm{\mu}$m, and
$a_{10} = 5\,420.08~\mathrm{\mu}$m. Thus, the spatial density distribution for a given dust
particle $i$ is

\begin{equation}
    \rho_{\mathrm{d}_i}=\frac{4\pi}{3}\rho_\mathrm{g} n_0 \int_{a_{\mathrm{edge}_i}}^{a_{\mathrm{edge}_{i+1}}}a^{\alpha+3} \,\mathrm{d}a
\end{equation}
where $a_{\mathrm{edge}_i}$ and $a_{\mathrm{edge}_{i+1}}$ are the edges of the bin. In consequence, 
each simulation cell is filled with a total dust density 
\begin{equation}
    \rho_\mathrm{d} = \sum_{i=1}^{10}\rho_{d_i}
\end{equation}
that varies with the
adopted size distribution and grain composition, as it is shown in Table~\ref{tab:rho_d}.

\begin{table}
\caption{Nominal spatial density distribution $\rho_\mathrm{d}$ (g\,cm$^{-3}$) for each dust
family and size distribution.}
  \label{tab:rho_d}
  \centering
  \begin{tabular}{lccc}
  \hline
    \hline
    $\alpha$ & Silicate & Carbonate & Iron \\
    \hline
    $-3.5$ & 1.5$\times 10^{-14}$ & 1.0$\times 10^{-14}$ & 3.6$\times10^{-14}$\\
    $-3.0$ & 6.0$\times 10^{-12}$ & 4.1$\times 10^{-12}$ & 1.4$\times10^{-11}$\\
    $-2.5$ & 3.0$\times 10^{-9}$ & 2.0$\times 10^{-9}$ & 7.2$\times 10^{-9}$\\
    $-2.0$ & 1.5$\times 10^{-6} $& 1.0$\times 10^{-6}$ & 3.6$\times 10^{-6}$\\
    \hline
    \end{tabular}
\end{table}

\subsection{Determination of dust temperature}\label{temperature}

In order to predict the polarized thermal emission of dust, the
dust temperature has to be set in \textsc{RADMC-3D}. One can either set a temperature
gradient \textit{ad-hoc}, or it can be computed with the \textit{mctherm} routine
available in \textsc{RADMC-3D}. This routine calculates dust temperature using the Monte Carlo method of \citet{Bjorkman2001} with various improvements such as the continuous absorption method of \citet{Lucy1999}.
Our main source of photons is the Sun, that is included in the model in the form
of a stellar spectrum\footnote{The solar spectrum has been downloaded from the CALSPEC data base (\url{https://www.stsci.edu/hst/instrumentation/reference-data-for-calibration-and-tools/astronomical-catalogs/calspec}) \citep{Bohlin2014} and has been extended to microwaves as the tail of a blackbody at 5\,780~K.}.
\begin{figure*}
    \centering
    \includegraphics[width=.90\textwidth]{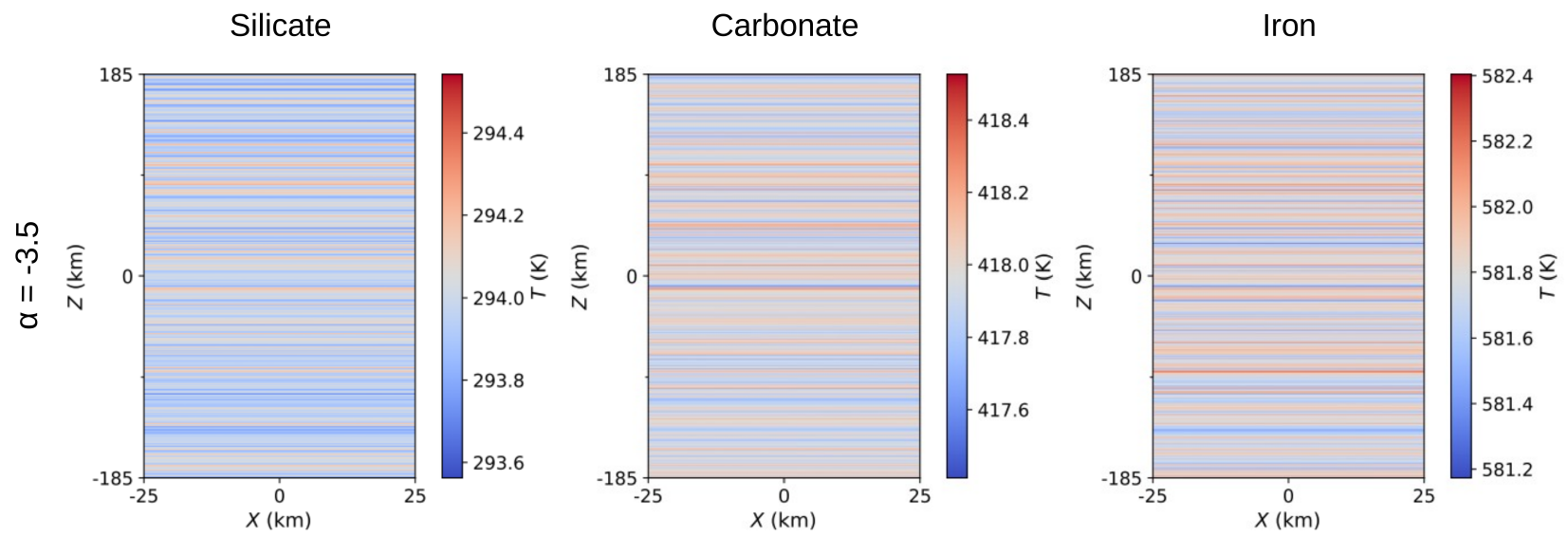}
    \includegraphics[width=.90\textwidth]{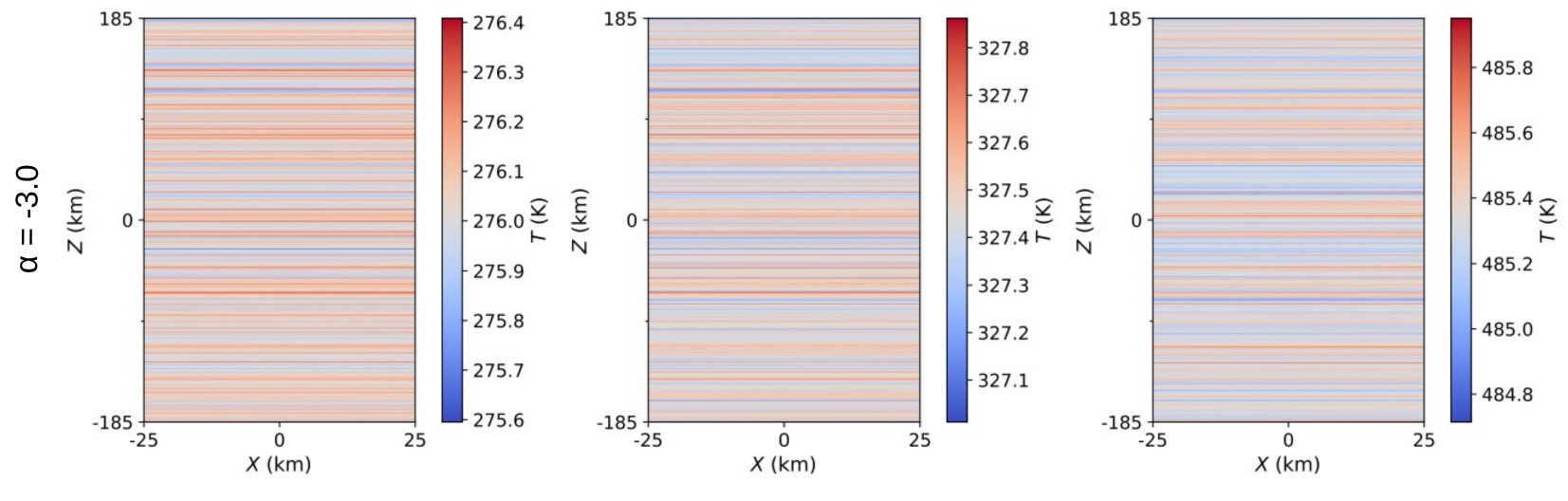}
    \includegraphics[width=.90\textwidth]{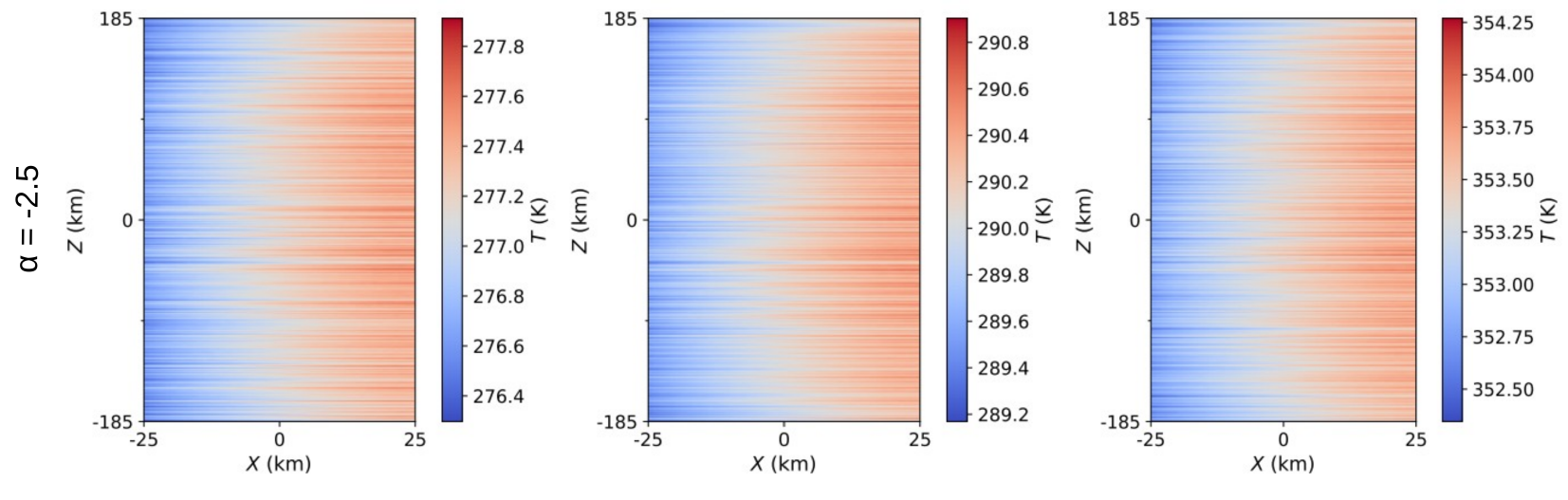}
    \includegraphics[width=.90\textwidth]{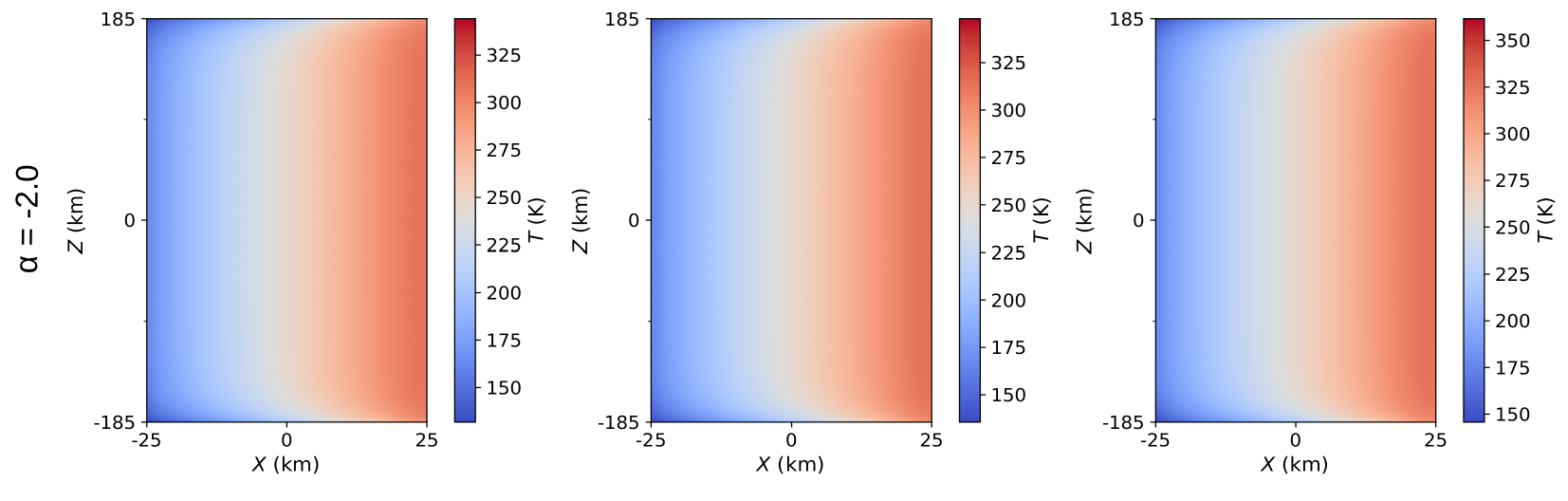}
    \caption{Temperature structures within the considered grid for different values of $\alpha$ from an altitude of 130~km to 500~km above Earth surface.}
    \label{fig:temperature_distribution}
\end{figure*}
For the temperature calculation, the total stellar luminosity is binned into a discrete number of photon packages, $nphot$, and 
setting $nphot$ = 5$\times 10^{10}$ in our models
proved to be a good compromise between computational cost and smooth temperature maps. 
The dust temperature is computed after a series of absorption, emission and scattering events 
following Eq.~6 in \citet{Bjorkman2001},
assuming that dust is in local thermodynamical equilibrium and the dust
temperature of a given cell is:
\begin{equation}
\sigma T^{4} = \frac{NL}{4N_{\gamma}\sum_{l=1}^{10}\kappa_{\rm P}^{l}(T)m_{l}}
\label{eq:dust_temperature}
\end{equation}
where $N$ is the number of photon packages absorbed by the cell, $L$
is the total luminosity, $N_{\gamma}$ is the total number of photon packages
(or $nphot$, as called before), $\kappa_{\rm P}$ is the Planck mean opacity and 
depends on the dust opacities, $m$ is the dust mass inside the cell, and 
the sum is taken over the 10 dust species considered in this work. This approach provides fair
results for optically-thin media, but for optically-thick configurations
($\alpha = -2.0$ models)
the results might be slightly biased.

In Fig.~\ref{fig:temperature_distribution}, the predicted temperature distribution 
is displayed for the three grain compositions and $\alpha$ values used in this work; 
these values match well the expectations for the solar radiation field
\citep[see][Fig.~24.4]{Draine_book}, where astrosilicates are expected to reach temperatures around 300~K whereas carbonaceous can heat up between 300 and 400~K. 
Remember that the $X$ axis points toward the Sun, so the population located at positive $X$ axis on the grid receives firstly the solar radiation independently on the satellite position around the orbit. The denser the medium, the stronger is the temperature gradient on the grid. Notice that the pronounced gradient of temperatures that is produced along the $X$ axis (for instance, in the models with $\alpha=-2.0$ or $-2.5$) is an unrealistic effect of defining an isolated and limited dust cloud with a bounded grid. For the less dense models, the temperature maps show statistical fluctuations in magnitude  due to computational reasons, but the temperature fluctuations are really small (of the order of $\sim$1~K).
We also find a dependency on composition, since carbonate grains reach higher temperatures than silicates, but
iron grains are the ones that reach the highest one. 
Another finding is tightly related to the fact that
the smaller-sized population 
clearly deviates from the expected equilibrium
temperature in all cases. As a test case, we computed the temperature
maps for all dust sizes separately (populations involved in Eq.~\ref{eq:dust_temperature})
and for the three materials and the value of $\alpha=-2.5$. We found that 
0.09~$\mathrm{\mu}$m graphite grains reach $T= 630$~K for this model; iron grains smaller than $a = 21.58$~$\mathrm{\mu}$m would 
exceed 460~K, while the smallest (0.02~$\mathrm{\mu}$m) particles reach
a temperature of around 840~K. However, since we are imposing thermal
equilibrium of all dust populations inside a given computational cell, these
small particles bias the temperature toward higher values. In order
to quantify the uncertainties introduced, we have computed the dust temperature
but considering only populations $l = 3, ..., 10$ for $\alpha = -3.5$
in Eq.~\ref{eq:dust_temperature}
for graphite and iron grains, which are the most extreme cases since these are
the models with more small-sized dust particles. We obtain values that are 
$\sim 0.85 T_{\rm dust}^{\rm all}$, the dust temperature considering the full population;
therefore, we expect that the polarization signal will be overpredicted by
a factor of at most 15 per cent.

\subsection{Polarization maps}
Finally, once we have the dust temperature maps it is possible to compute the
polarization emission. 
In this case, the declination of the satellite
will make a difference for those cases where $B$-alignment of dust is possible; from
the discussion on Sect.~\ref{alignment}, we know that irons can show 
both $B$- and $k$-alignment depending on grain size, while
graphites and silicates can only be aligned with the radiation field. Hence, in
order to have the two extreme cases, we generate the polarization maps
for two different points of the orbit, the South magnetic pole and the equator, to
have the two extreme magnetic field orientations (parallel and perpendicular to
the line of sight, respectively). For the time being, grain alignment has
to be set by the user in \textsc{RADMC-3D}, and only one alignment direction for each
simulation
is allowed; therefore, we have assumed perfect alignment for each
particle, that depending on grain material and size will be $B$- or $k$-alignment
as stated before. In consequence, we obtain in most cases $k$-alignment maps
for the whole dust population, although polarization from the smaller
iron grains is split in two maps ($B$- and $k$-aligned as required).

Apart from setting the alignment of dust particles, to produce
a polarized signal we need to include nonspherical grains. 
Following the standard implementation in \textsc{RADMC-3D}, 
this can be done by assigning
different weights to the projection of the opacity along the two directions perpendicular to the line of sight, 

\begin{subequations}
    \begin{align}
         k_{\nu,h}(\psi) &= 1 \\
         k_{\nu,v}(\psi) &= (1-A\cos(\pi\cos\psi))/(1+A)
    \end{align}
    \label{kcoef}
\end{subequations}
where $\psi$ is the angle between the grain minor axis and the line of sight and $A$ is a factor related to the axial ratio of the oblate spheroid $s = a_1 : a_2$ (with $a_1$ and $a_2$, the semi-minor and semi-major axis, respectively). In this work, 
the value of $A$ is fixed to 0.5, emulating an aspect ratio of 1:3
(oblate grains) that is a common value used for fitting the interstellar 
extinction \citep{Das2010}.

Polarization is computed in terms of the Stokes vector ($I$, $Q$, $U$, $V$). Following the IAU 1974 definition \citep{IAUStokes}, we choose a coordinate system ($X$, $Y$, $Z$) such that the value of $Q$ is positive along the $X$ axis. An image of the expected polarized emission can be computed using the {\it image} routine of \textsc{RADMC-3D}.

Scattering can also be included in \textsc{RADMC-3D}, however,
for our models the optical depth $\tau$ is less than 1 (and therefore scattering is negligible), except for the silicate model with $\alpha = -2.0$, where $\tau \sim$ 700, meaning optically thick medium, and scattering might be significant.
Neglecting scattering renders make impossible to introduce a phase shift in the thermal
polarized signal of the oblate aligned grains, so we cannot have circular polarization
and the $V$ component of the Stokes vector is null.

\section{Results}\label{results}

We have obtained $I$ and $Q$ maps for all the considered species
(silicates, graphites and irons), the four values of the slope of the
size distribution $(\alpha = -3.5, -3.0, -2.5, -2.0)$, and two extreme
points of the orbit (South magnetic pole and equator).
Besides, according to the calculations presented in Sect.~\ref{alignment},
graphites will always be $k$-aligned, silicates will also be $k$-aligned
but for the smallest (0.02~$\mathrm{\mu}$m) population (and do not present
a relevant emission, so will be ignored), and irons will be the ones
to present both $B$-alignment for sizes smaller than 4~$\mathrm{\mu}$m
(populations $a_{1}$--$a_{4}$, remember Fig.~\ref{fig:t_B/t_flight})
and $k$-alignment for the larger-sized populations. 

\begin{figure*}
    \centering
    \includegraphics[width=.95\textwidth]{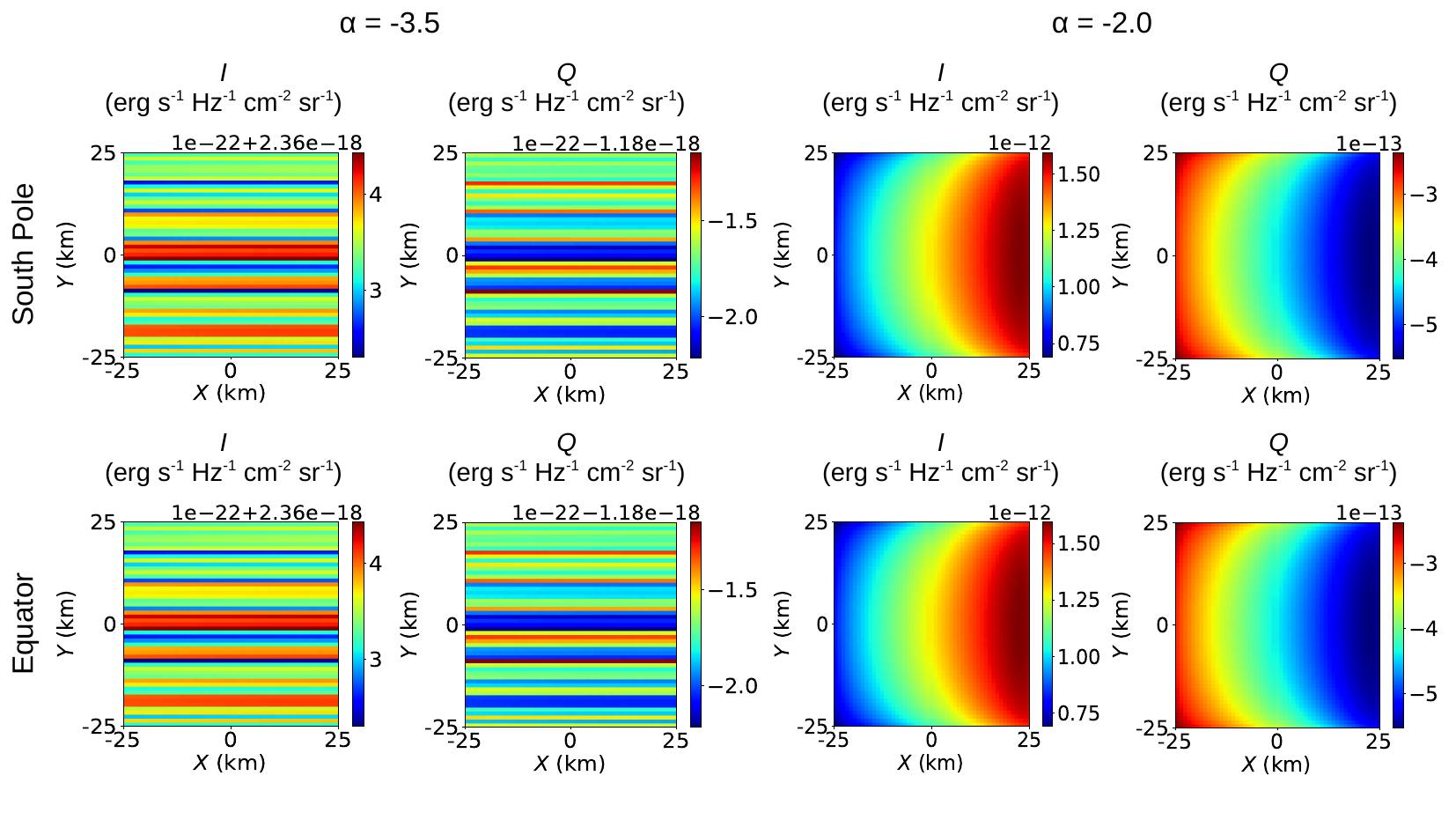}
    \caption{$I$ and $Q$ Stokes parameters for
    silicate grains measured at 220~GHz at the South magnetic pole (top)
    and at the equator (bottom) aligned with respect to the 
    solar radiation field ($k$-alignment). We note that maps at both
    values of $\alpha$ reproduce the spurious patterns seen in the temperature
    maps in Fig.~\ref{fig:temperature_distribution}, being the ones
    in the $\alpha = -2.0$ due to higher opacities and boundary effects, and 
    the strike patterns seen in the $\alpha = -3.5$ maps only due to
    statistical fluctuations in the spatial generation of the photon packages, 
    resulting in negligible spatial variations in the $I$ and $Q$ magnitudes.
    Maps for the South magnetic pole are identical to those
    of the equator, since both configurations are equivalent. Negative values
    of $Q$ are due to the linear polarization along the $Y$ axis of the image. 
    }
    \label{fig:kalignment_IQ_silicates}
\end{figure*}

In all cases, $k$-alignment $(I, Q)$ maps present the same behavior for a given
$\alpha$: they do not depend on the position along the orbit
and grain material only influence the magnitude of the Stokes vectors.
As an illustration, we present
in Fig.~\ref{fig:kalignment_IQ_silicates} the maps at 220~GHz for silicates
and the size distributions with 
$\alpha = -2.0$ and $\alpha = -3.5$. It can be seen that $(I, Q)$ 
maps are identical for the South magnetic pole and the equator, since
the relative alignment of dust particles with respect to the field of view
of the satellite is identical when dust grains are aligned with respect
to the solar radiation. In addition, we see a strong dependency of the 
magnitude of the Stokes vector on the dust size distribution, since 
there are six orders of magnitude of difference in signal between the
$\alpha = -2.0$ and $\alpha = -3.5$ maps.\par

\begin{figure*}
    \centering
    \includegraphics[width=.95\textwidth]{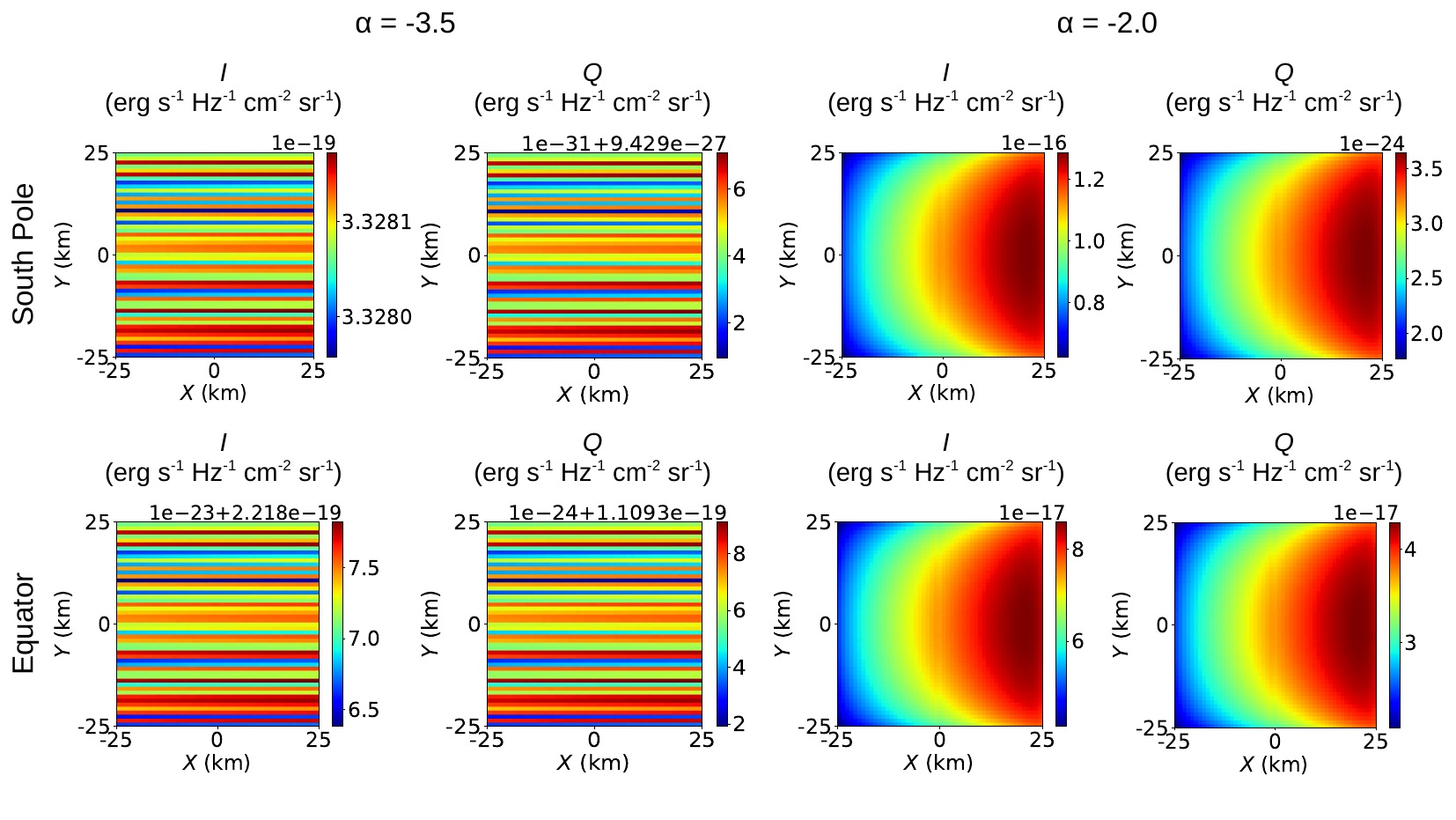}
    \caption{Same as Fig.~\ref{fig:kalignment_IQ_silicates}, but
    for iron grains aligned with respect to the Earth's
    magnetic field ($B$-alignment). Note the differences in magnitude
    between the South magnetic pole and equator maps because of the dependence on 
    the observing point. In this case, positive values
    of $Q$ are due to the linear polarization along the $X$ axis of the image.}
    \label{fig:Balignment_IQ_irons}
\end{figure*}

On the other hand, $B$-alignment maps will do depend on the vantage
point of the telescope. Since irons are the ones that may present an
efficient alignment with respect to the Earth magnetic field, in Fig.~\ref{fig:Balignment_IQ_irons} 
we present the equivalent $(I, Q)$ maps
for irons at 220~GHz at the same slopes as for
Fig.~\ref{fig:kalignment_IQ_silicates}. In this case, we do observe
differences in signal between the South magnetic pole and the equator:
the magnitude of $Q$ is seven (eight) orders of magnitude higher
in the equator than in the South pole for  $\alpha = -2.0$ ($\alpha = -3.5$).
This is because dust grains are oriented with
their semi-minor axis parallel to the magnetic field; therefore, the projected cross sections viewed from the South magnetic pole by the satellite are perfect spheres 
and the polarization is negligible. On the contrary, in the equator we see spheroids along the line of sight, resulting in a higher polarization for this population.

\begin{figure*}
    \centering
    \includegraphics[width=.95\textwidth]{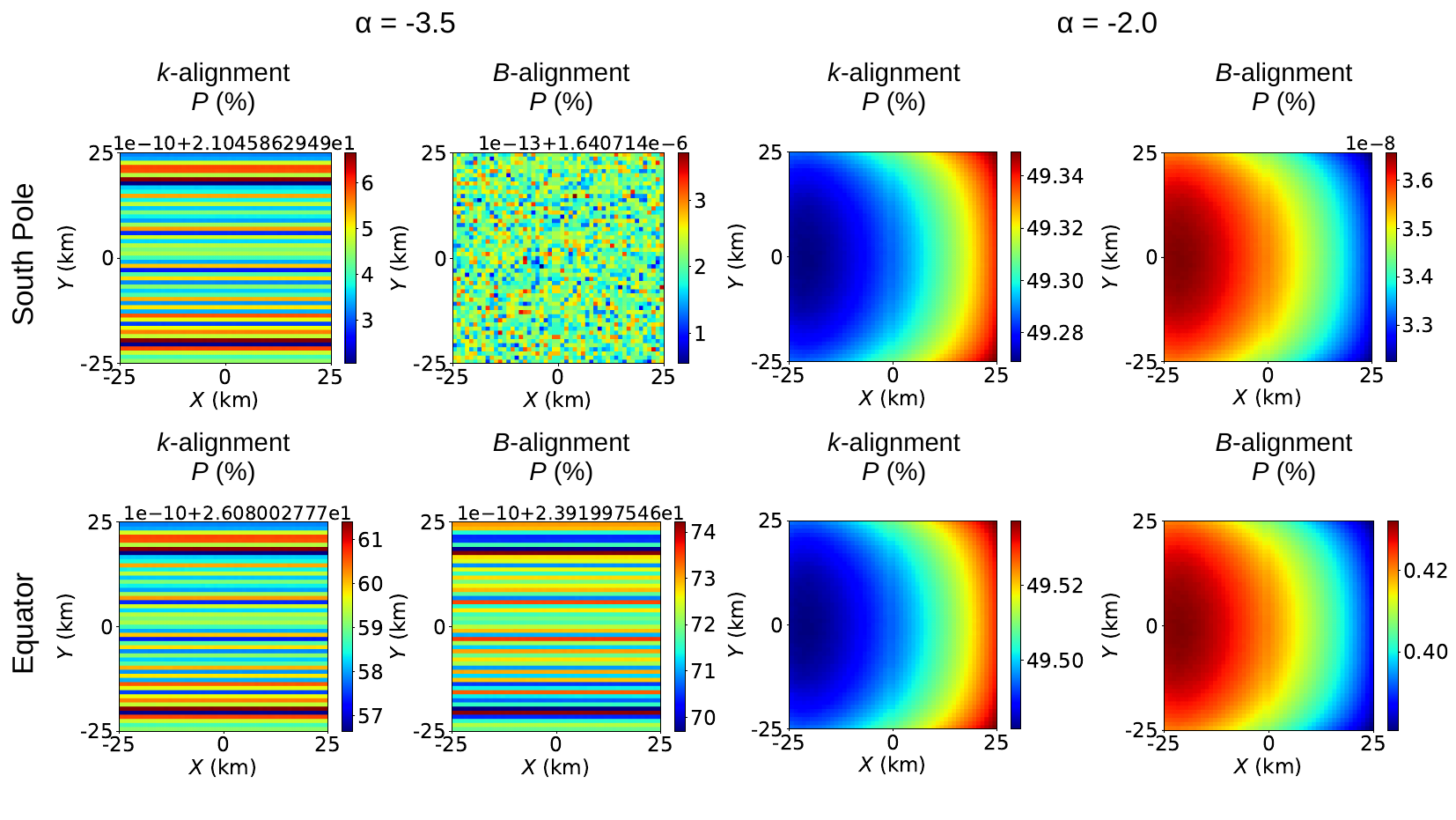}
    \caption{Polarization maps for iron grains for the two extremes of the
    considered size distributions ($\alpha = -3.5$ and $\alpha = -2.0$). Polarization is calculated following Eq.~\ref{Pdust}.}
    \label{fig:iron_polarization}
\end{figure*}

As an illustration, in Fig.~\ref{fig:iron_polarization} we present the 
polarization maps for iron grains for both alignments, the two slopes
and the two magnetic field orientations, computed
as 
\begin{equation}
    P_{\rm al}(\%) = \frac{Q_{\rm al}}{(I_k+I_B)}\times 100
    \label{Pdust}
\end{equation}
where the subscript $al$ refers to the considered alignment ($k$ or $B$). 
For $k$-alignment maps, the morphology
for silicates and graphites are equivalent, and there is not any significant
differential polarization inside the field of view depending on the vantage point. However, for the $B$-alignment maps we do observe strong variations in polarization, since
in the South pole we will barely have signal from irons.

\begin{figure}
    \centering
    \includegraphics[width=88 mm]{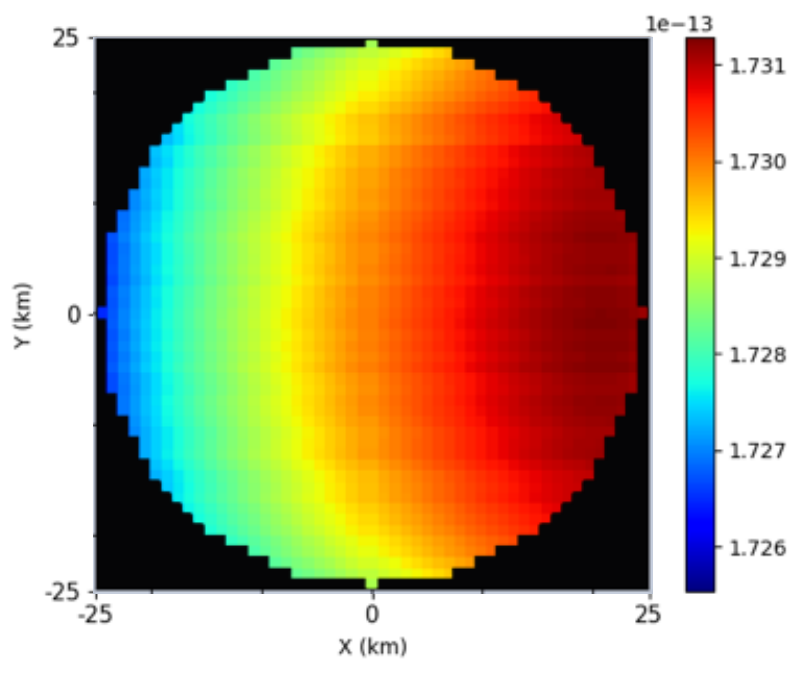}
    \caption{Example of the simulated MARTINLARA field of view of the intensity
    map for $k$-aligned
    silicates at 220~GHz and with $\alpha = -2.5$. This spot is the one
    used for the computation of the mean values of the polarization vector 
    summarized in Tables~\ref{tab:table_silicates}, \ref{tab:table_carbonates},
    and \ref{tab:table_iron}.}
    \label{fig:masked_image}
\end{figure}

Finally, to provide a quantitative discussion of the results, we
have computed the mean value of intensity $<I>$, the magnitude $<Q>$, and percent polarization
$<P>$ for all our simulations; these averages are computed inside
MARTINLARA consortium's circular field of view, so only part of the maps have been
used (see Fig.~\ref{fig:masked_image} for an illustration). The results are 
shown in Tables~\ref{tab:table_silicates}, \ref{tab:table_carbonates}, 
and \ref{tab:table_iron} for silicates, graphites, and irons respectively.
From these tables, we can see that the mean dust intensity at
220~GHz is higher than the signal at 80~GHz for all cases. 
However, for some densities, we observe that the percent polarization at $k$-alignment is more reduced for 220~GHz in comparison to 80~GHz. 
Besides, for $k$-aligned grains there is a clear relationship between grain composition
and mean intensity: from lower to higher signal we have irons, 
graphites and silicates, but the magnitudes are constant along the orbit.
The outcome radiation of the grains is the result of various processes represented in the radiative transfer equation where both the grain optical properties and grain temperature take part of. Although iron grains heat up the most, they are always the weakest in signal because they present the lowest efficiencies of the optical properties at microwave range.
In addition, there is a strong dependency on the size distribution: the shallower
the slope (lower $|\alpha|$), the stronger the dust emission, and the larger
the differences between materials; for $\alpha = -3.5$, the average intensity
of silicates is larger by a factor of $\sim$1.6--2.8 of those of
graphites, depending on the 
frequency considered, but for $\alpha = -2.0$ the differences can be of the
order of $\sim 20$. In practice, a shallow dust size distribution implies
a higher abundance of large dust grains, so we can conclude from this that
dust emission at the frequencies considered in this study
will be dominated by large dust grains. 
Moreover, comparing the relative mean intensities of
silicates and the rest of the species it is clear that they will dominate 
the emission, while the contribution
of irons to the total signal will have a lower impact on a standard
size distribution with $\alpha = -3.5$, and will be negligible for higher
values of $\alpha$.\par 

With respect to the contribution of $B$-aligned grains, in the view
of the results for irons shown in Table~\ref{tab:table_iron}, we will focus on the 
signal at the equator, which is the only region where we can expect to
detect the polarized emission. We see that while the average intensity
for $k$- and $B$-aligned dust particles is similar for $\alpha = -3.5$ (although
it is still higher for the latter by a factor of 1.6), for shallower
distributions where the relative number of small ($B$-aligned) dust
grains begins to be scarce the total contribution to the emissivity
is very low. Therefore, in terms of total average intensity we can say that
irons play a modest role for interstellar-like dust size distributions 
($\alpha = -3.5$), but as a general rule
the signal will be dominated by large silicate grains.

\begin{landscape}
\begin{table}
\caption{Average intensity and polarization fractions for the simulations
of silicates.}
  \label{tab:table_silicates}
  \centering
  \footnotesize
  \begin{tabular}{lccccccc}
    \hline \hline
    \multicolumn{8}{c}{$k$-alignment}\\
    \hline
    & & \multicolumn{3}{c}{80~GHz}&\multicolumn{3}{c}{220~GHz}\\
    & & $\langle I \rangle$& $\langle Q \rangle$ 
    & $\langle P \rangle$ 
    & $\langle I \rangle$ & $\langle Q \rangle$ 
    & $\langle P \rangle$ \\
    & & (erg\,s$^{-1}$\,Hz$^{-1}$\,cm$^{-2}$\,sr$^{-1}$) & (erg\,s$^{-1}$\,Hz$^{-1}$\,cm$^{-2}$\,sr$^{-1}$)
    & \%
    & (erg\,s$^{-1}$\,Hz$^{-1}$\,cm$^{-2}$\,sr$^{-1}$) & (erg\,s$^{-1}$\,Hz$^{-1}$\,cm$^{-2}$\,sr$^{-1}$)
    &\% \\
    \hline
$\alpha=-3.5$ & South pole & 8.61$\times10^{-20}$ & -4.31$\times10^{-20}$
& 49.99  
& 2.36$\times10^{-18}$ & -1.18$\times10^{-18}$
& 50.00 \\
& Equator & 8.61$\times10^{-20}$ & -4.31$\times10^{-20}$
& 49.99  
& 2.36$\times10^{-18}$ & -1.18$\times10^{-18}$
& 50.00 \\
$\alpha=-3.0$ & South pole & 3.31$\times10^{-17}$ & -1.65$\times10^{-17}$
& 49.99 
& 5.51$\times10^{-16}$ & -2.76$\times10^{-16}$
& 49.99 \\
& Equator & 3.31$\times10^{-17}$ & -1.65$\times10^{-17}$
& 49.99 
& 5.51$\times10^{-16}$ & -2.76$\times10^{-16}$
& 49.99 \\
$\alpha=-2.5$ & South pole & 1.41$\times10^{-14}$ & -6.96$\times10^{-15}$
& 49.48 
& 1.73$\times10^{-13}$ & -8.50$\times10^{-14}$
& 49.15 \\
& Equator & 1.41$\times10^{-14}$ & -6.96$\times10^{-15}$
& 49.48 
& 1.73$\times10^{-13}$ & -8.50$\times10^{-14}$
& 49.15 \\
$\alpha=-2.0$ & South pole & 9.02$\times10^{-14}$ & -3.79$\times10^{-14}$
& 41.99 
& 1.25$\times10^{-12}$ & -4.28$\times10^{-13}$
& 34.28 \\
& Equator & 9.02$\times10^{-14}$ & -3.79$\times10^{-14}$
& 41.99 
& 1.25$\times10^{-12}$ & -4.28$\times10^{-13}$
& 34.28 \\
    \hline    
    \end{tabular}
\end{table}
\end{landscape}

\begin{landscape}
\begin{table}
\caption{Same as Table~\ref{tab:table_silicates} but for graphites.}
  \label{tab:table_carbonates}
  \centering
  \footnotesize
  \begin{tabular}{lccccccc}
    \hline \hline
    \multicolumn{8}{c}{$k$-alignment}\\
    \hline
    & & \multicolumn{3}{c}{80~GHz}&\multicolumn{3}{c}{220~GHz}\\
    & & $\langle I \rangle$& $\langle Q \rangle$ 
    & $\langle P \rangle$ 
    & $\langle I\rangle$ & $\langle Q \rangle$ 
    & $\langle P \rangle$ \\
    & & (erg\,s$^{-1}$\,Hz$^{-1}$\,cm$^{-2}$\,sr$^{-1}$) & (erg\,s$^{-1}$\,Hz$^{-1}$\,cm$^{-2}$\,sr$^{-1}$)
    & \% 
    & (erg\,s$^{-1}$\,Hz$^{-1}$\,cm$^{-2}$\,sr$^{-1}$) & (erg\,s$^{-1}$\,Hz$^{-1}$\,cm$^{-2}$\,sr$^{-1}$)
    &\%  \\
    \hline
$\alpha=-3.5$ & South pole &  5.40$\times10^{-20}$ & -2.70$\times10^{-20}$
& 50.00  
& 8.42$\times10^{-19}$ & -4.21$\times10^{-19}$
& 50.00  \\
& Equator & 5.40$\times10^{-20}$ & -2.70$\times10^{-20}$
& 50.00 
& 8.42$\times10^{-19}$ & -4.21$\times10^{-19}$
& 50.00  \\
$\alpha=-3.0$ & South pole & 4.15$\times10^{-18}$ & -2.07$\times10^{-18}$
& 49.99 
& 5.11$\times10^{-17}$ & -2.55$\times10^{-17}$
& 49.99  \\
& Equator & 4.15$\times10^{-18}$ & -2.07$\times10^{-18}$
& 49.99 
& 5.11$\times10^{-17}$ & -2.55$\times10^{-17}$
& 49.99 \\
$\alpha=-2.5$ & South pole & 7.91$\times10^{-16}$ & -3.95$\times10^{-16}$
& 49.97 
& 8.69$\times10^{-15}$ & -4.34$\times10^{-15}$
& 49.96 \\
& Equator & 7.91$\times10^{-16}$ & -3.95$\times10^{-16}$
& 49.97 
& 8.69$\times10^{-15}$ & -4.34$\times10^{-15}$
& 49.96 \\
$\alpha=-2.0$ & South pole & 4.56$\times10^{-15}$ & -2.26$\times10^{-15}$  
& 49.64 
& 5.56$\times10^{-14}$ & -2.75$\times10^{-14}$
& 49.41 \\
& Equator & 4.56$\times10^{-15}$ & -2.26$\times10^{-15}$
& 49.64 
& 5.56$\times10^{-14}$ & -2.75$\times10^{-14}$
& 49.41 \\
    \hline    
    \end{tabular}
\end{table}  
\end{landscape}

\begin{landscape}
\begin{table}
\caption{Same as Table~\ref{tab:table_silicates} but for irons.}
\label{tab:table_iron}
 \centering
 \footnotesize
  \begin{tabular}{lccccccc}
    \hline \hline
    \multicolumn{8}{c}{$k$-alignment}\\
    \hline
    & & \multicolumn{3}{c}{80~GHz}&\multicolumn{3}{c}{220~GHz}\\
    & & $\langle I \rangle$& $\langle Q \rangle$ 
    & $\langle P\rangle$ 
    & $\langle I \rangle$ & $\langle Q \rangle$ 
    & $\langle P \rangle$ \\
    & & (erg\,s$^{-1}$\,Hz$^{-1}$\,cm$^{-2}$\,sr$^{-1}$) & (erg\,s$^{-1}$\,Hz$^{-1}$\,cm$^{-2}$\,sr$^{-1}$)
    & \%
    & (erg\,s$^{-1}$\,Hz$^{-1}$\,cm$^{-2}$\,sr$^{-1}$) & (erg\,s$^{-1}$\,Hz$^{-1}$\,cm$^{-2}$\,sr$^{-1}$)
    &\%  \\
    \hline
$\alpha=-3.5$ & South pole &  1.87$\times10^{-20}$ & -9.36$\times10^{-21}$
& 25.98
& 2.42$\times10^{-19}$ & -1.21$\times10^{-19}$
& 21.05 \\
& Equator &  1.87$\times10^{-20}$ & -9.36$\times10^{-21}$
& 30.94
& 2.42$\times10^{-19}$ & -1.21$\times10^{-19}$
& 26.08 \\
$\alpha=-3.0$ & South pole & 9.96$\times10^{-19}$ & -4.98$\times10^{-19}$
& 44.11
& 1.23$\times10^{-17}$ & -6.17$\times10^{-18}$
& 41.86\\
& Equator & 9.96$\times10^{-19}$ & -4.98$\times10^{-19}$
& 45.91
& 1.23$\times10^{-17}$ & -6.17$\times10^{-18}$
& 44.26\\
$\alpha=-2.5$ & South pole &  1.34$\times10^{-16}$ & -6.72$\times10^{-17}$ 
& 49.67
& 1.56$\times10^{-15}$ & -7.81$\times10^{-16}$
& 49.51\\
& Equator & 1.34$\times10^{-16}$ & -6.72$\times10^{-17}$
& 49.78 
& 1.56$\times10^{-15}$ & -7.81$\times10^{-16}$
& 49.67 \\
$\alpha=-2.0$ & South pole & 6.62$\times10^{-16}$ & -3.31$\times10^{-16}$ 
& 49.48
& 8.57$\times10^{-15}$ & -4.28$\times10^{-15}$ 
& 49.29\\
& Equator & 6.62$\times10^{-16}$ & -3.31$\times10^{-16}$ 
& 49.63 
& 8.57$\times10^{-15}$ & -4.28$\times10^{-15}$ 
& 49.50\\
    \hline \hline
    \multicolumn{8}{c}{$B$-alignment}\\
    \hline
    & & \multicolumn{3}{c}{80~GHz}&\multicolumn{3}{c}{220~GHz}\\
    & & $\langle I \rangle$& $\langle Q \rangle$ 
    & $\langle P\rangle$ 
    & $\langle I \rangle$ & $\langle Q \rangle$ 
    & $\langle P \rangle$ \\
    & & (erg\,s$^{-1}$\,Hz$^{-1}$\,cm$^{-2}$\,sr$^{-1}$) & (erg\,s$^{-1}$\,Hz$^{-1}$\,cm$^{-2}$\,sr$^{-1}$)
    & \%
    & (erg\,s$^{-1}$\,Hz$^{-1}$\,cm$^{-2}$\,sr$^{-1}$) & (erg\,s$^{-1}$\,Hz$^{-1}$\,cm$^{-2}$\,sr$^{-1}$)
    &\% \\
    \hline
$\alpha=-3.5$ & South pole & 1.73$\times10^{-20}$ & 4.90$\times10^{-28}$
& Negligible
& 3.33$\times10^{-19}$ & 9.43$\times10^{-27}$
& Negligible\\
&Equator &  1.15$\times10^{-20}$ & 5.77$\times10^{-21}$
& 19.06 
& 2.22$\times10^{-19}$ & 1.11$\times10^{-19}$
& 23.92\\ 
$\alpha=-3.0$ & South pole &  1.33$\times10^{-19}$ & 3.77$\times10^{-27}$
& Negligible 
& 2.40$\times10^{-18}$ & 6.79$\times10^{-26}$
& Negligible \\
&Equator &  8.86$\times10^{-20}$ & 4.43$\times10^{-20}$
& 4.09 
& 1.60$\times10^{-18}$ & 7.99$\times10^{-19}$
& 5.74 \\
$\alpha=-2.5$ & South pole &  8.80$\times10^{-19}$ & 2.49$\times10^{-26}$
& Negligible 
& 1.53$\times10^{-17}$ & 4.34$\times10^{-25}$
& Negligible \\
&Equator &  5.86$\times10^{-19}$ & 2.93$\times10^{-19}$
& 0.22
& 1.02$\times10^{-17}$ & 5.10$\times10^{-18}$
& 0.32\\
$\alpha=-2.0$ & South pole &  6.29$\times10^{-18}$ & 1.78$\times10^{-25}$
& Negligible 
& 1.07$\times10^{-16}$ & 3.04$\times10^{-24}$
& Negligible\\
&Equator &  4.19$\times10^{-18}$ & 2.10$\times10^{-18}$
& 0.32
& 7.16$\times10^{-17}$ & 3.58$\times10^{-17}$
& 0.42\\
    \hline    
    \end{tabular}
\end{table} 
\end{landscape}

Due to the geometry of our system, the sign of the $Q$ magnitude of the Stokes vector, meaning, the direction of the linear polarization, will reveal the alignment direction that is produced: a negative $Q$ is related to a $k$-alignment, exhibiting a linear polarization in the $Y$ direction of the image. On the other hand, positive $Q$ means that $B$-alignment is occurring with the linear polarization in the $X$ axis of the image. Therefore, the two types of considered alignment produces linear polarization perpendicular to each other. In both cases, the direction of polarization is perpendicular to the minor axis of the grains pointing in the same direction as the corresponding alignment field.\par

Regarding the dust polarization, we can see for a $k$-alignment that for silicates and carbonate, and for almost all size distribution, the percentage is around 50 per cent for any point along the orbit at any of the observing frequencies, except for the silicates with $\alpha=-2.0$ that deviate to values ranging from 42 per cent (80~GHz) to 34 per cent (220~GHz). For these frequencies, silicate is the only material 
which absorption opacities are comparable in magnitude to the scattering ones; moreover, the higher density of the model
with $\alpha=-2.0$ might also produce the difference in $P$ for that case. 
As we have already inferred from Fig.~\ref{fig:iron_polarization} for $B$-aligned grains, $P$ at the South pole is extremely low compared to the signal received at the equator that ranges between $\sim$24 per cent
to 0.32 per cent in polarization for irons depending on size distributions and frequency.
$B$-aligned iron polarization is weaker than $k$-aligned ones because the smallest iron grains, that are $B$-aligned, have lower optical efficiencies at the frequencies we are working on.

\begin{figure*}
    \centering
    \includegraphics[width=.85\textwidth]{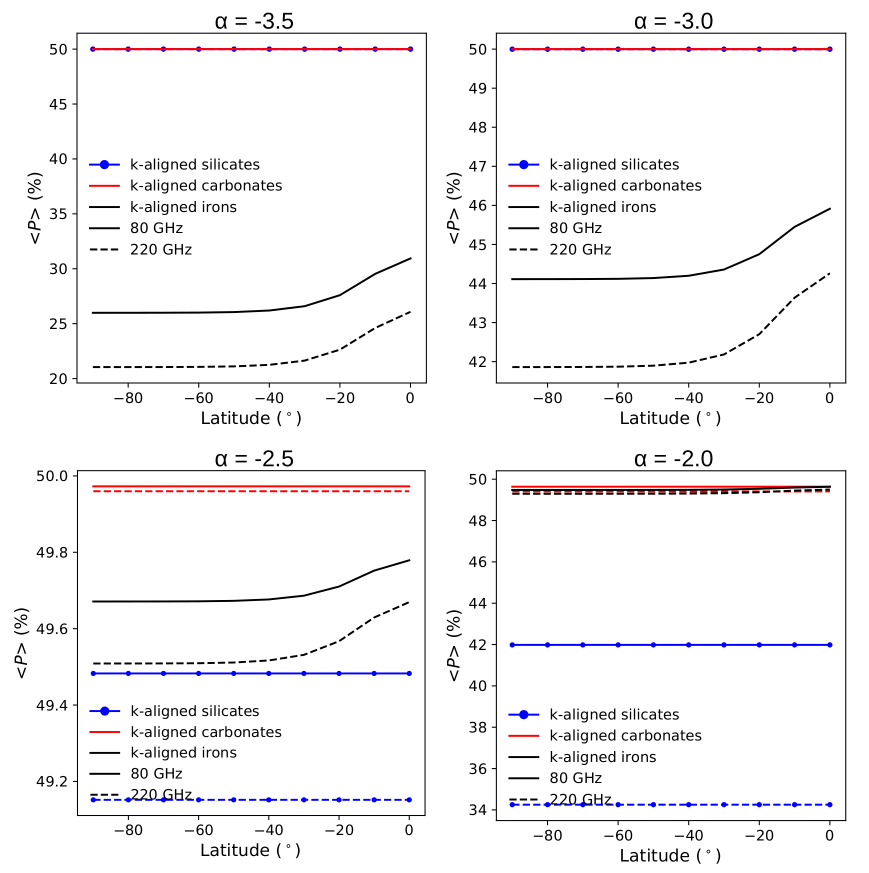}
    \caption{Percent polarization along the orbit for the three dust species with $k$-alignment and all size distributions at 80 and 220~GHz.}
    \label{fig:P_lat_k}
\end{figure*}

\begin{figure}
    \centering
    \includegraphics[width= 88mm]{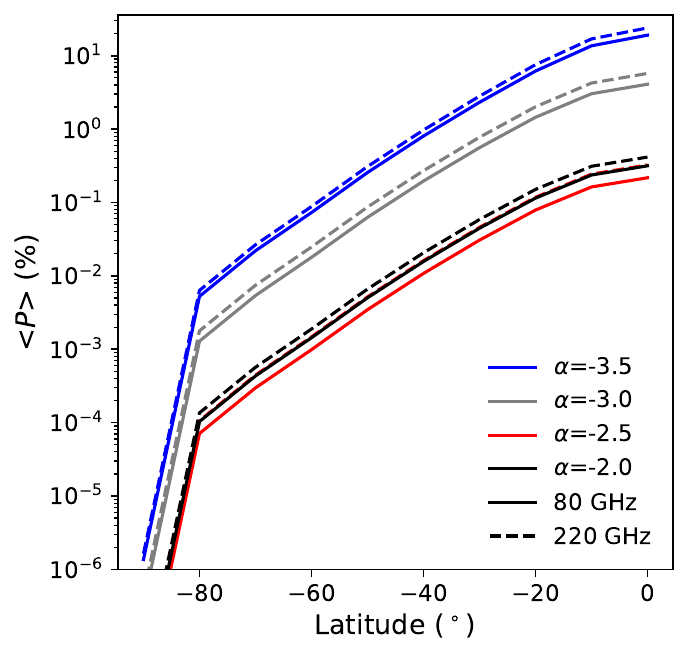}
    \caption{Percent polarization along the orbit for the $B$-aligned irons and all size distributions at 80 and 220~GHz. A cutoff at $10^{-6}$ is performed, considering the below values imperceptible.}
    \label{fig:P_lat_B}
\end{figure}

The variations along the orbit of the percentage of polarization for $k$-alignment are shown in Fig.~\ref{fig:P_lat_k}. We can see that $P$ is constant for silicates and carbonates. On the other hand, the percentage of polarization of irons is modulated by their total emission, composed of the signal of the grains with both alignments. Since the unpolarized signal at equator is lower than that produced at the South pole, the percentage of polarization suffers a slight increase at South pole for irons. While increasing $\alpha$, emission from $B$-aligned grains become weaker than the produced by $k$-aligned dust, therefore this rise reduces.
In Fig.~\ref{fig:P_lat_B} tendency of the polarization for $B$-alignment irons is shown. We can see that the polarization is governed by the orientation of the grain with the observation point, being the equator the favorable point.

\section{Discussion}\label{discussion}
\subsection{Evaluation of the impact of the Earth thermosphere in grain alignment}

The propagation of the streams of space dust within the rarefied environment of the Earth thermosphere may also affect the alignment of the dust grains.
The physical problem has been studied in the context of the alignment of dust grains with a supersonic gas flow \citep{Gold1952a, Gold1952b} or even in subsonic \citep{LH07b} named mechanical torque (MET) alignment in the context of interstellar medium research. 
For the current investigation, the inertial system is set in the moving dust stream, and the standing gas in the thermosphere 
is treated as the high velocity flow with typical speed in the range between 11.2 and 71 km~s$^{-1}$ (see Sect.~\ref{alignment}). 
For the subsequent analysis, we will use the models put forward by these authors but caution that the theory is strongly dependent
on the properties of the grain surface on which the mechanical torques act (geometry, mechanical characteristics of the materials involved, etc) 
and that more theoretical and observational studies are required to mature the theory \citep{HL2018}; see, {\it e.g.} the recent analysis by 
\citet{Reissl2023} of the efficiencies of spin up and precession. Note also, that this analysis assumes that there is not a gas flow associated to 
the meteoritic showers but just particles and rocks.

Following the procedure described in Sect.~\ref{alignment} for the analysis of the relevance of the alignment torques, we have computed the 
the timescale for the mechanical alignment of the dust grains and compare it with the flight time and the time scales of the other alignment 
mechanisms addressed in this work.

Since the streaming velocity of the meteoritic dust is supersonic, the mechanical Gold alignment should be considered first. The alignment timescale, $\tau_{Gold}$ is given by  \citep{Hoang&Lazarian2013},

\begin{equation}
    \tau_{Gold}= \frac{16\rho_g k_B T_{\rm gas}a}{15 \gamma_{\rm flow}^2 \sum_n (n_{\rm n}m_{\rm n}^2)v^3}
\end{equation}
with $\gamma_{\rm flow}$ the anisotropy of the flow.
\begin{figure}
    \centering
    \includegraphics[width=88mm]{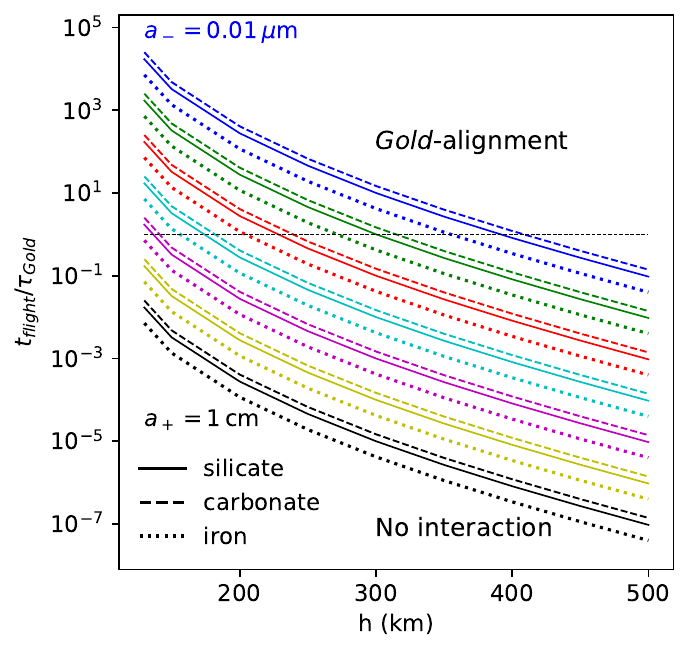}
    \caption{Timescale of Gold alignment compared to the flight time of the meteoritic dust.}
    \label{fig:tflight_tgold}
\end{figure}
This alignment is found to be efficient when compared with the flight time only for the smallest grains and when the thermosphere becomes denser, as we can see in Fig.~\ref{fig:tflight_tgold}. In addition, we consider the mechanical torques proposed by \citet{LH07b}. According with the theory, this timescale, $t_{MET}$ is:

\begin{equation}
    t_{MET} = \frac{2\pi I \omega}{\pi a^3 \sum_n (n_{\rm n} m_{\rm n}) v^2 Q_{\rm prec}}
\end{equation}
being $Q_{\rm prec}$ the efficiency factor of the torque, taken equal to 0.1 following \citet{Hoang2022}.
In Fig.~\ref{fig:tv_tflight} we can appreciate the high efficiency of the mechanical torques. The MET alignment timescale is significantly shorter than the
flight time for all grains larger than 1 $\rm \mu$ above 130 km. Let us now, evaluate whether MET alignment is more efficient than RAT alignment.
As shown in Fig.~\ref{fig:tv_tk}, MET is very efficient and the dominant alignment mechanism in the layers of the thermosphere.

\begin{figure}
    \centering
    \includegraphics[width=88 mm]{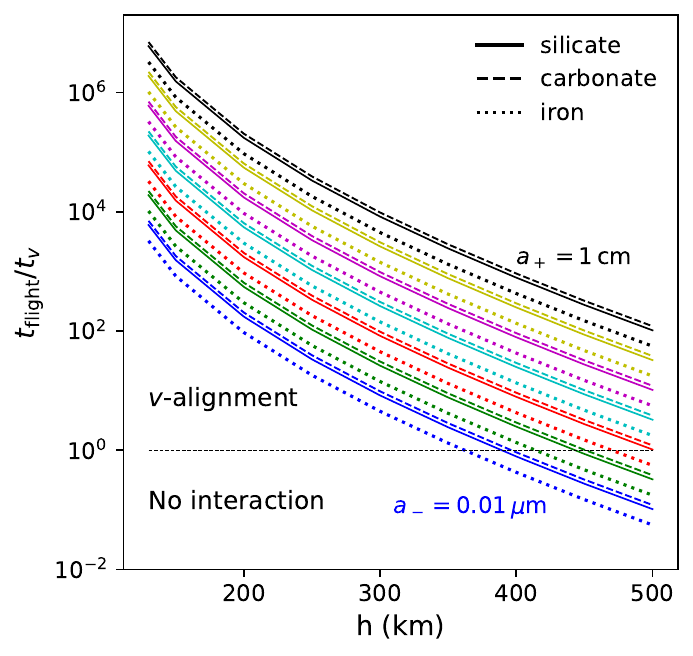}
    \caption{Ratio between the timescale for dust alignment by mechanical torques and the time flight. The collisions with the thermospheric gas suffice for the alignment of the 
    grains larger than 1 $\mu$m above 130.}
    \label{fig:tv_tflight}
\end{figure}

\begin{figure}
    \centering
    \includegraphics[width=88 mm]{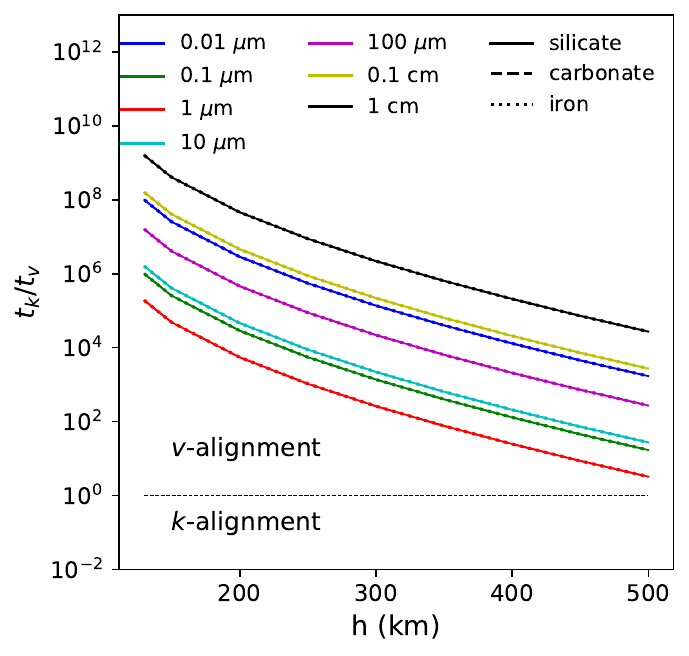}
    \caption{Ratio between the timescale of mechanical and radiative torques. MET alignment is more efficient than RAT alignment for all the grain sizes
    considered in this work.Also, note that the ratio between MET and RAT is independent on the grain material.}
    \label{fig:tv_tk}
\end{figure}

\begin{figure}
    \centering
    \includegraphics[width=88 mm]{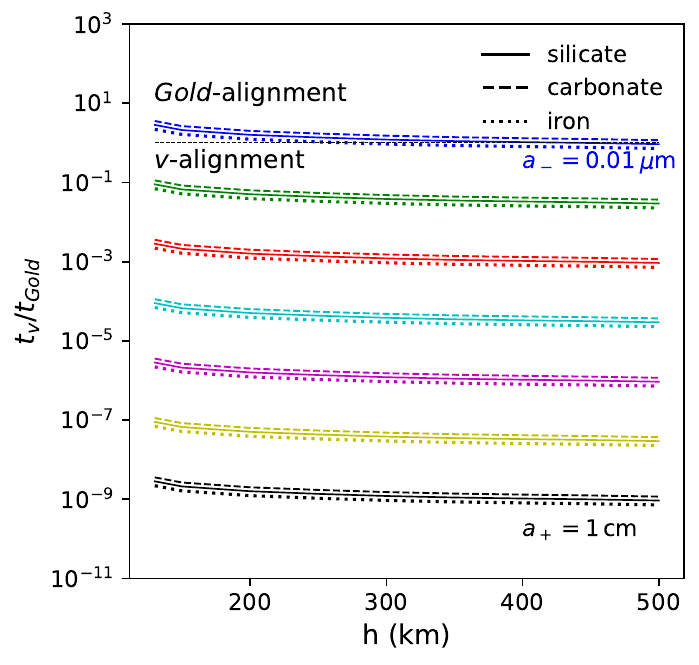}
    \caption{Ratio of MET timescale over Gold-alignment time for the species considered in this work. Among alignment caused by grain-gas interaction, we see that alignment via METs is more efficient for all materials and sizes except for the smallest dust grains.}
    \label{fig:Gold-MET}
\end{figure}

For completeness, in Fig.~\ref{fig:Gold-MET} we have compared both mechanisms of alignment caused by grain-gas interactions, Gold-alignment and METs, and we have seen that the efficiency of METs is higher than Gold-alignment unless for the smallest grains of 0.01 $\rm \mu$m at specific altitudes. Since the contribution to the signal of only this size population at microwave frequencies might be negligible, we dismiss Gold-alignment.

In summary, MET-alignment might be the dominant mechanism at the atmospheric layers of interest. We have calculated the expected signal of the grain population with this alignment.

We have considered two extreme directions for simulating the trajectory that falling grains follow and, in consequence, their velocity vector. We have established that grains enter with an angle of 30$^{\rm o}$ respect to the ground in the plane $XZ$ and also in $YZ$ plane of our grid.

\begin{figure*}
    \centering
    \includegraphics[width=.95\textwidth]{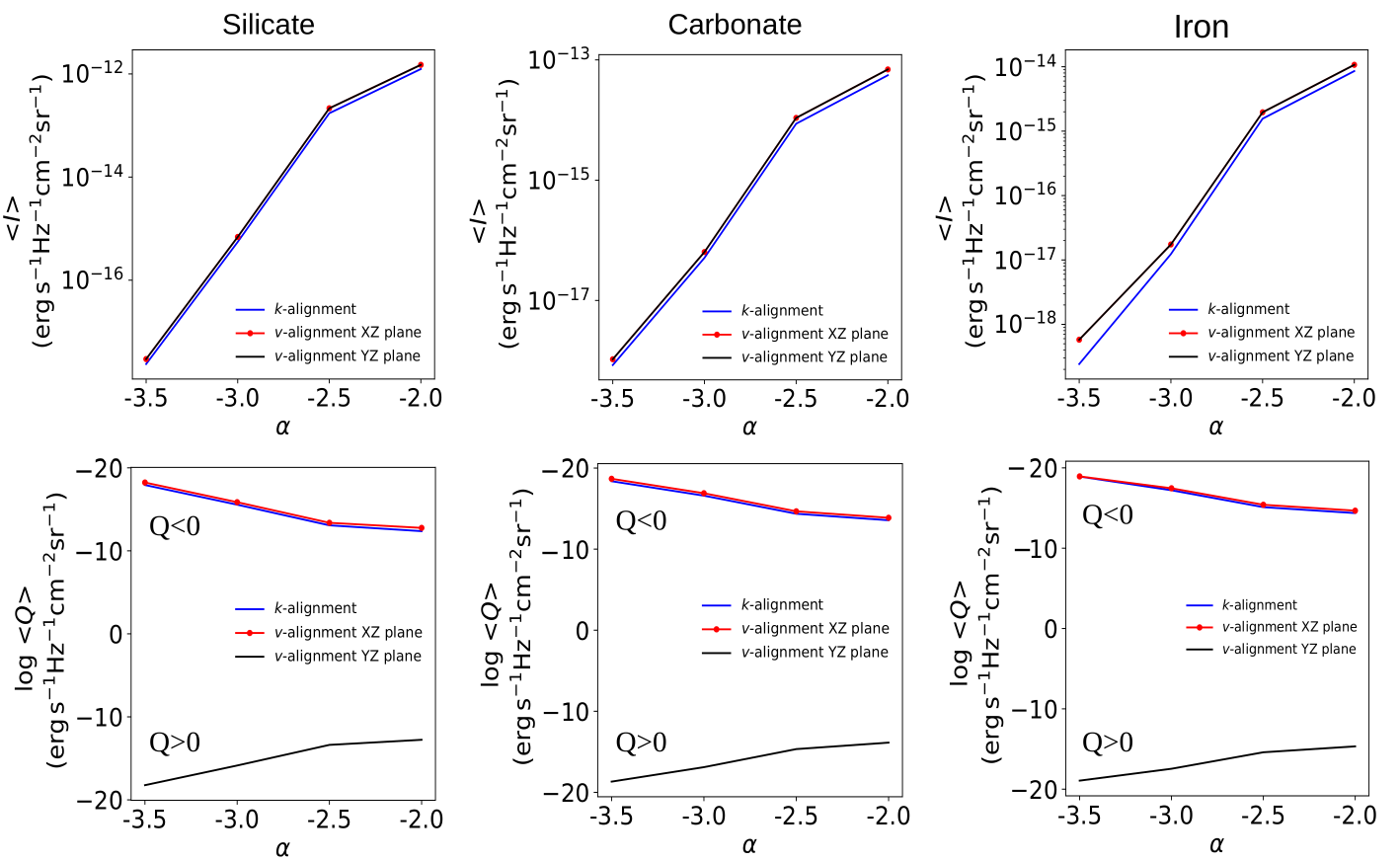}
    \caption{Mean intensity and $Q$ parameter values for silicates, carbonates and irons when they are $v$-aligned. Grain velocity of 11.2 km\,s$^{-1}$ and frequency of 220 GHZ have been taken. We also plot the same values for $k$-alignment as a reference.}
    \label{fig:IQ_MET}
\end{figure*}

From the simulations carried out, we see that the pattern in the synthetic image is the same as in the case of $k$-alignment. In Fig.~\ref{fig:IQ_MET}, we represent the mean values of $I$ and $Q$ for the three materials following $v$-alignment. The intensity behaviour is the same as in the $k$-alignment scenario and we see that is about 1.25 times greater than the intensity for the $k$-alignment, unless for the case of irons with $\alpha=$-3.5 where the ratio rises to 2.4, measured at 220 GHz. Regarding to the polarization, we appreciate a strong dependence between $Q$ and the direction of the trajectory, even producing a change of sign depending on the falling plane. The signal of polarization of $v$-alignment is about a half of the expected for $k$-alignment, except for irons with $\alpha=$-3.5 where is near the same (0.95 times $Q_k$). These variations in the two modes of alignment are the result of the obliquity of the grains with the line of sight.

Direction of linear polarization $Q$ strongly depends on the direction of the falling trajectory of the grains. Each meteoroid and dust cloud have different trajectories depending on their parent body. Collecting several polarization signals over time, if polarization varies significantly over time and changes sometimes its sign, we can suspect that a MET alignment is happening. Thus, regarding cosmic dust falling to Earth from space can offer an scenario for MET alignment theory proof.
\subsection{Detectability}
The main motivation of this work is to assess the detectability of the
polarization signal of infalling dust particles into the Earth's atmosphere
by a small technological probe operating from LEO at microwave wavelengths.

For this purpose, basic mission and instrumentation constraints
have been advanced by the MARTINLARA consortium, which has developed
novel photonics microwave radiometers that can operate without cryogenic cooling. Laboratory tests carried by the team
provided some basic constraints such as the power detection threshold (8~erg\,s$^{-1}$) for operation in the 80--220~GHz band with bandwidth 1~GHz,
and collecting surface of 10~cm$^2$.  

Let us go back to the predictions summarized in Tables~\ref{tab:table_silicates}--\ref{tab:table_iron}. Because silicates are the dust population more abundant, we take as an example the intensity and polarization predicted
for silicate grains space shower with a size distribution parametrized with  $\alpha = -2.0$ at 220~GHz.
According to the results in Table~\ref{tab:table_silicates}, the expected intensity of the radiation at 500~km height (LEO orbit) is $1.25 \times 10^{-12}$~erg\,s$^{-1}$\,Hz$^{-1}$\,cm$^{-2}$\,sr$^{-1}$ per square kilometer (grid element). 
This accounts for a total power of 24.5~erg\,s$^{-1}$ reaching the detector since the antenna beam is circular with a projected spot radius of 25~km.
Therefore, the signal that is produced is high enough to be detected over the sensitivity threshold.

\begin{figure}
    \centering
    \includegraphics[width=88 mm]{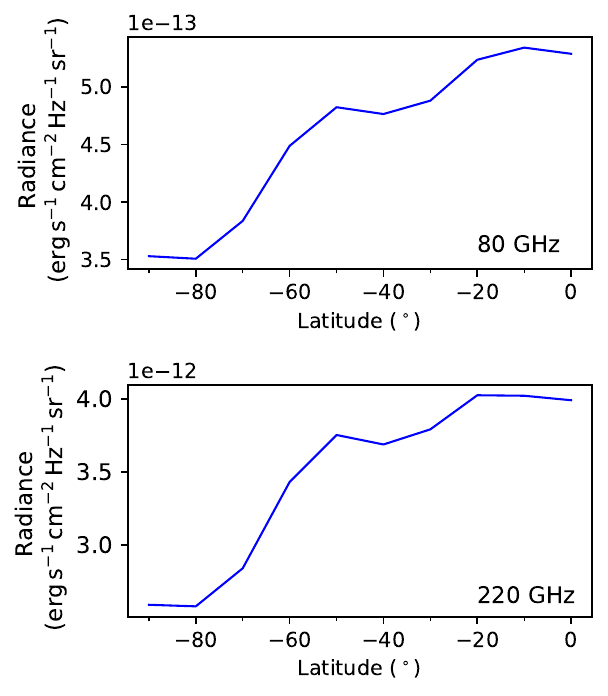}
    \caption{Earth radiance as a function of latitude for 80~GHz (top) and 220~GHz (bottom) predicted by the
    Planetary Spectrum Generator.}
    \label{fig:earth_radiance}
\end{figure}

Only 34.28 per cent of this radiation is polarized, which accounts for a total power of 8.4~erg\,s$^{-1}$ of polarized radiation that
has to be discriminated from the background unpolarized component of the radiation of the dust but also from the strong microwave background produced by the Earth thermal emission. 
The expected flux can be calculated using the Planetary
Spectrum Generator (PSG) \footnote{PSG can be used from \url{ https://psg.gsfc.nasa.gov/index.php}} \citep{Villanueva2018, Villanueva2022}.
Inputs for the calculation are observing view-point, which has been set to a  circular,
dawn-dusk  LEO orbit at 500~km over the Earth surface, with tentative inclination $\sim -85\degr$
to flight over the Earth South magnetic pole. The radiance has been computed for two baseline frequencies,
80~GHz and 220~GHz, and it is represented in Fig.~\ref{fig:earth_radiance}, for the relevant range of latitudes.
As expected, the Earth background radiation at microwave wavelengths lowers by a factor of $\sim 2$ from the equator to the pole facilitating the detection at very high latitudes.

The detector conceived by the MARTINLARA consortium is designed to operate with two antennas/channels (H and V) and to detect signals linearly polarized in two perpendicular planes. 
The accuracy of the antennas in the selection of the position angle of the polarized radiation
has been estimated to be $\sim 0.1 ^\mathrm{o}$ hence, only a small fraction of the unpolarized radiation is actually detected (0.1/180). Let us assume that
the H channel is oriented parallel to the polarization plane of the radiation produced by the dust and the V channel is perpendicular to it.
The rate between the powers received by both antennas would be given by

\begin{equation}
   \frac{H}{V} = \frac{P_{\rm pol} + P_{back}}{P_{\rm back}} 
\end{equation}
where $P_{\rm pol} = 8.4$~erg\,s$^{-1}$ and $P_{\rm back}$ is the total background radiation power that for the observation at 220~GHz of a cloud located
over the Earth equator is 0.053~erg\,s$^{-1}$, 83 per cent caused by the Earth background and the rest from unpolarized dust radiation.
Thus,

\begin{equation}
    \frac{H}{V} = \frac{8.4+0.053}{0.053} = 160.
\end{equation}

Laboratory tests have shown that simple (though very sensitive) instrument such as the designed by MARTINLARA can discriminate the
rate of the power received by both channels, H/V,  to levels as high as 32 dB (32dB = 1585); this accuracy is more than enough to measure
accurately the polarization predicted from the simulations for this silicate dust cloud with $\alpha = -2.0$.

Obviously, more sensitive receivers and higher collecting surfaces are required to detect graphites or irons. For instance, if we considered only a 1\,m telescope, still smaller than the instrument of the Planck mission, we would be able to reach intensities of three orders of magnitude highers than with our proposal. However, the potentials of a small
probe for global studies of space dust infalling on Earth should not be neglected.

\begin{figure}
    \centering
    \includegraphics[width=88 mm]{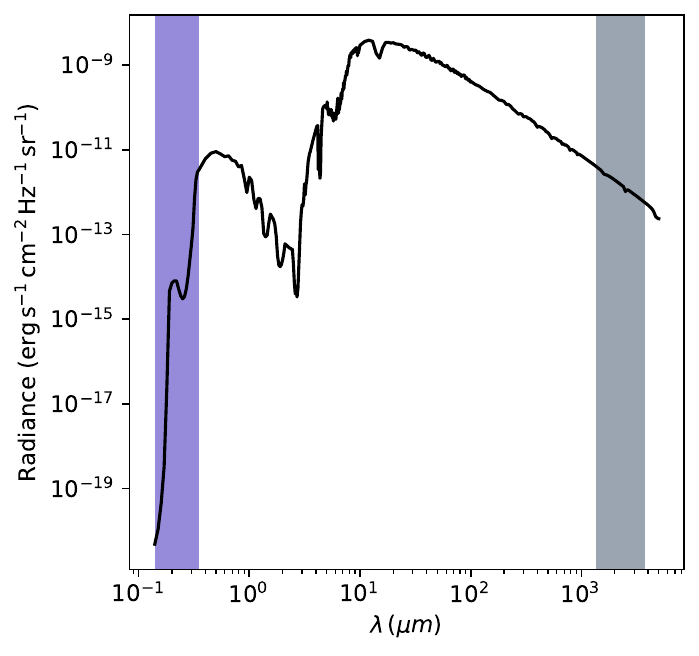}
    \caption{Earth radiance as a function of wavelength. The grey 
    shadowed area represents the microwave range considered
    in this work, and the
    violet area is the ultraviolet range from 0.14 to 0.35~$\mathrm{\mu}$m 
    where the Earth background is lower. }
    \label{fig:earth_radiance_wav}
\end{figure}

Also should be noted that the Earth background drops by two orders of magnitude at ultraviolet wavelengths, as shown in  Fig.~\ref{fig:earth_radiance_wav}
\citep{GomezdeCastro2023}. An evaluation of the detectability of space dust at this wavelength is deferred for a later work.

\section{Conclusions}\label{conclusions}
This paper presents the first estimates of the detectability
of infalling interplanetary dust particles into the Earth atmosphere from space (from LEO orbit)
at microwave wavelengths (80--220~GHz) 
with high sensitivity radiometers that do not require cryogenic cooling. 
For this work, we have considered three materials, silicates, graphites and irons, which
are commonly present in comets and asteroids. We have estimated the polarized
emission of these species with the radiative transfer Monte Carlo code
\textsc{RADMC-3D}, considering oblate dust particles aligned with the solar
radiation field (silicates, graphites and 
irons with sizes larger than 4~$\mathrm{\mu}$m) and with the Earth magnetic field 
(small irons). We have restricted our study to heights that go from
the satellite orbit (500~km) down to 130~km, that is the limit
below which any grain alignment is lost due to collisions with the 
dense atmosphere; no meteoroid fragmentation or ablation are expected 
in this range. We have built a grid of models that covers several
size distributions with power law indices $\alpha = (-3.5, -3.0, -2.5, -2.0)$,
assuming a constant particle number of 0.22~cm$^{-3}$. Our results can be summarized
as follows:

\begin{enumerate}
\item Thermal emission for silicate grains is the highest of the three considered materials, followed by carbonates and, finally, iron grains are the ones that present the lowest signal at the microwave frequencies we work with.
\item The polarized signal of infalling interplanetary dust particles 
varies with dust size, composition, and size distribution index $\alpha$. 
\item We can retrieve the type of alignment that the grains present ($k$- or $B$-alignment) by looking the direction of $Q$ Stokes parameter and the behaviour of the percent polarization along the orbit.
\item Since only small iron particles interact efficiently with the 
Earth's magnetic field, and they produce a
characteristic
increase in the polarized
emission toward the equator not observed
for the rest of the materials, it would be possible
to use these curves to better constrain dust composition.

\item Due to the strong Earth background, only silicates, the most efficient dust grains,
would be detectable from space at microwave wavelengths.
\item Observation from space of material falling into Earth atmosphere can offer an opportunity of MET alignment theory verification collecting the variations of polarization signal over time.

\end{enumerate}

In this work, we have demonstrated that it is possible to detect
infalling material into the Earth before ablation occurs. A mission
such as MARTINLARA would provide complementary measurements for
ground-based observations that are limited to lower heights, and would
contribute to the characterization of interplanetary particles.
Besides, we also propose that another mission to measure
polarization at ultraviolet wavelengths would be useful since
the contribution of the Earth background at these
wavelengths is smaller.

\section*{Acknowledgements}
This work was funded by Comunidad de Madrid S2018/NMT-4333 MARTINLARA-CM and PID2020-116726RB-I00. 
L.B.-A. acknowledges Universidad Complutense de Madrid and Banco Santander
for a grant 'Periodo de Orientación Postdoctoral', and also the receipt 
of a Margarita Salas postdoctoral
fellowship from Universidad Complutense de Madrid (CT31/21), funded by
the 'Ministerio de Universidades' with Next Generation EU funds, that
supported her at different stages of this work.

\section*{Data Availability}
The data underlying this article will be shared on reasonable request to the corresponding author.

\bibliographystyle{mnras}
\bibliography{references}

\appendix

\section{Grain charge}\label{charge}
\begin{figure}
    \centering
    \includegraphics[width=88mm]{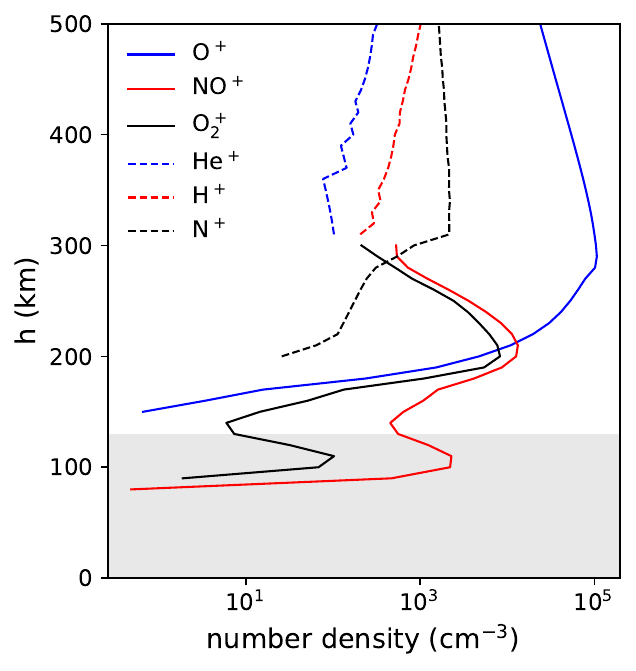}
    \includegraphics[width=88mm]{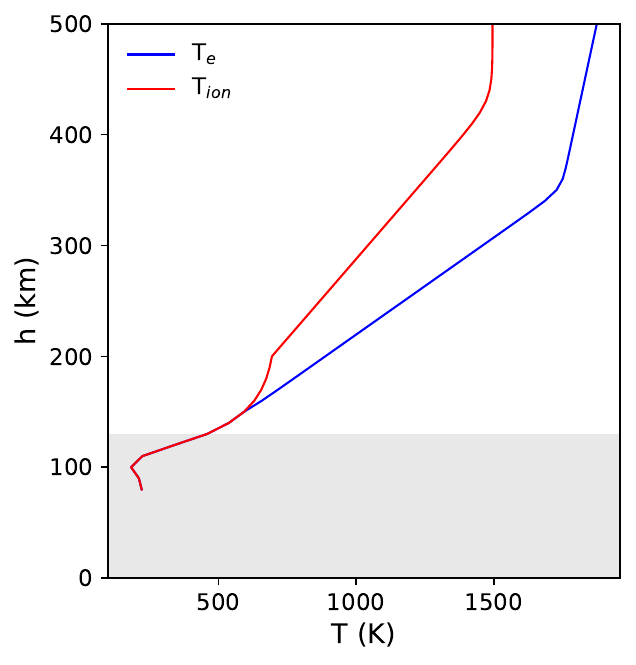}
    \caption{Number density of ions of the ionosphere ($top panel$) and temperature of the electrons ($T_e$) and ions ($T_{ions}$) ($bottom panel$). Data are extracted from IRI2016 model at South Pole on August 4, 2020.}
    \label{fig:ionosphere}
\end{figure}

Dust charge distribution has been computed from the equilibrium between photoionization and accretion of ions/electrons, using the freely available code \url{https://github.com/lbeitia/dust_charge_distribution} based on \citet{WD2001}. In this code, several ionic species can be introduced and it has the solar stellar spectra as an option. The input parameters of the gas atmosphere are the ones shown in Fig.~\ref{fig:ionosphere} obtained from 
IRI model \footnote{\url{https://kauai.ccmc.gsfc.nasa.gov/instantrun/iri/}}.

\begin{figure}
    \centering
    \includegraphics[width=58mm]{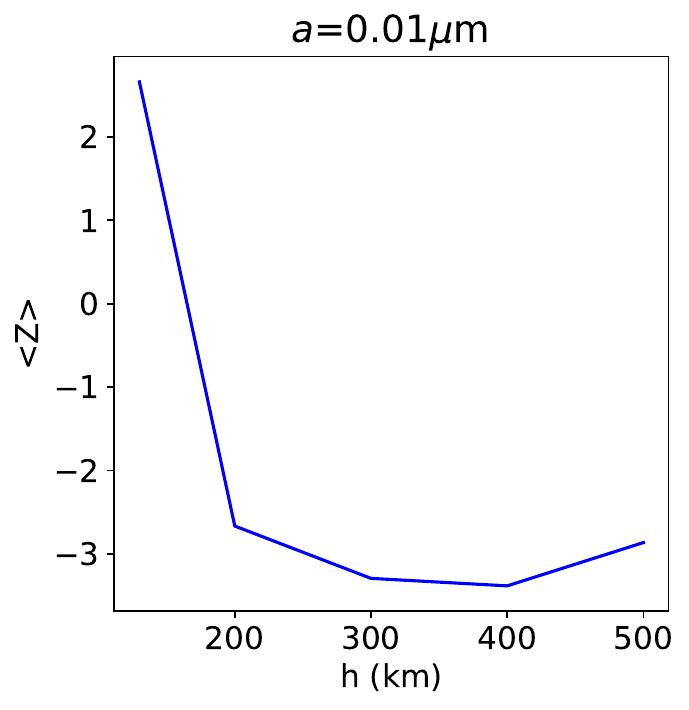}
    \includegraphics[width=58mm]{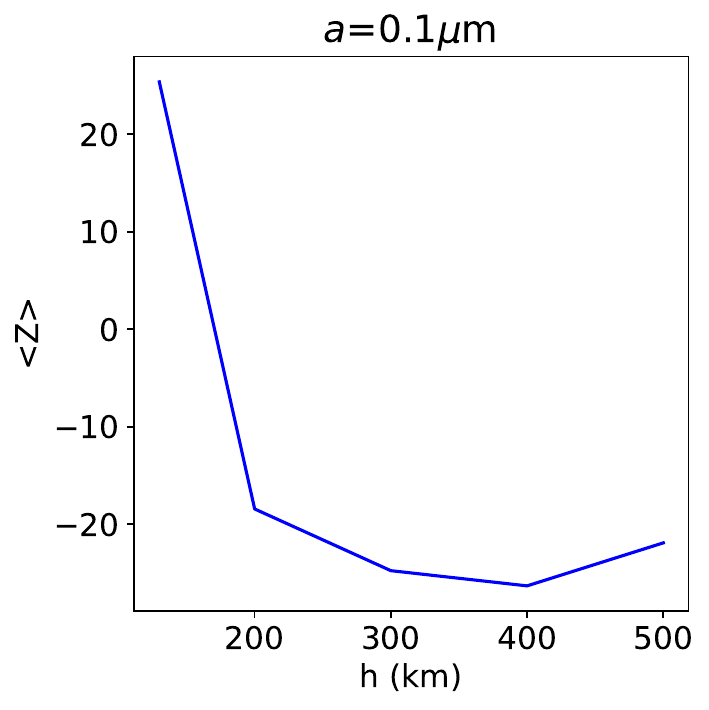}
    \includegraphics[width=58mm]{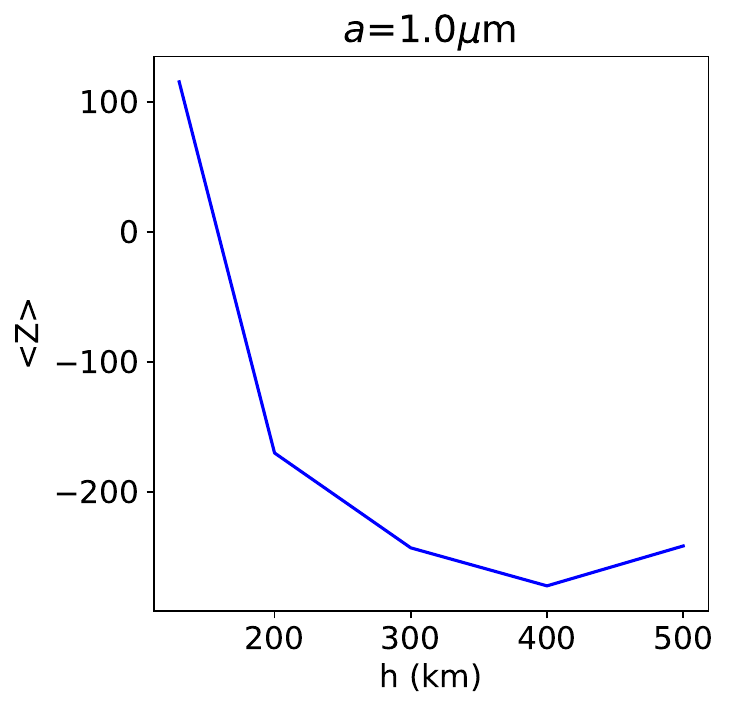}
    \includegraphics[width=58mm]{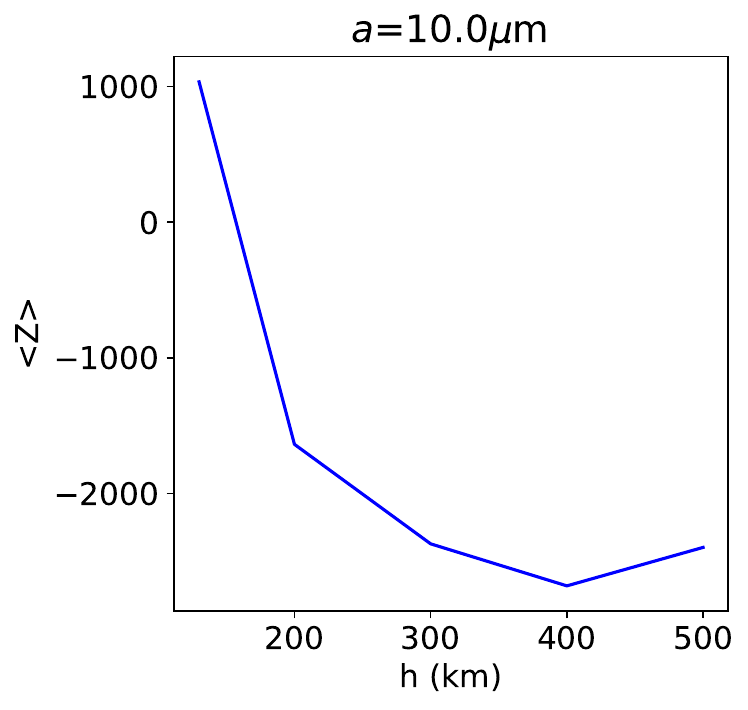}
    \caption{Mean dust charge for different grain sizes along the atmosphere altitude.}
    \label{fig:dust_charge}
\end{figure}

The mean dust charge are represented in Fig.~\ref{fig:dust_charge}, where we can see the variation of the charge of the particle at different heights of the atmosphere. We can calculate the electric moment of the grain:
\begin{equation}
    p = \epsilon eZa
\end{equation}
being $\epsilon=10^{-2}$ the parameter that describes the charge distribution and having into account that the charge distribution becomes dominant over the intrinsic dipole moment.

\bsp	
\label{lastpage}
\end{document}